\input harvmac
\overfullrule=0pt
\parindent 25pt
\tolerance=10000
\input epsf

\newcount\figno
\figno=0
\def\fig#1#2#3{
\par\begingroup\parindent=0pt\leftskip=1cm\rightskip=1cm\parindent=0pt
\baselineskip=11pt
\global\advance\figno by 1
\midinsert
\epsfxsize=#3
\centerline{\epsfbox{#2}}
\vskip 12pt
{\bf Fig.\ \the\figno: } #1\par
\endinsert\endgroup\par
}
\def\figlabel#1{\xdef#1{\the\figno}}
\def\encadremath#1{\vbox{\hrule\hbox{\vrule\kern8pt\vbox{\kern8pt
\hbox{$\displaystyle #1$}\kern8pt}
\kern8pt\vrule}\hrule}}

\font\cmss=cmss10
\font\cmsss=cmss10 at 7pt

\def\inbar{\vrule height1.5ex width.4pt depth0pt}

 \def\T{{\Theta}}
 \def\frac#1#2{{#1\over #2}}

 \def\s{\sqrt}

 \def\p{\partial}

 \def\al{\alpha'}
 \def\de{\partial}

 \def\f {\frac}
 \def\ti{\tilde}
 \def\ap{\alpha}

 \def\ddd{\cdot\cdot\cdot}
 
 \def\la{\langle}
 \def\lb{\rangle}

 \def\vp{\varphi}
\def\tvp{\ti{\varphi}}
\def\Str{{\rm Str}}
\def\T{{\tilde{T}}}
\def\p{\partial}
\def\str{{\rm Str}}

\def\IC{{\relax\,\hbox{$\inbar\kern-.3em{\rm C}$}}}
\def\IZ{\relax\ifmmode\mathchoice
{\hbox{\cmss Z\kern-.4em Z}}{\hbox{\cmss Z\kern-.4em Z}}
{\lower.9pt\hbox{\cmsss Z\kern-.4em Z}} {\lower1.2pt\hbox{\cmsss
Z\kern-.4em Z}}\else{\cmss Z\kern-.4em Z}\fi}
\def\IR{\relax{\rm I\kern-.18em R}}

\lref\V{
  C.~Vafa,
  ``Brane/anti-brane systems and U(N$|$M) supergroup,''
  arXiv:hep-th0101218.
}

\lref\AM{
  L.~Alvarez-Gaume and J.~L.~Manes,
  ``Supermatrix models,''
  Mod.\ Phys.\ Lett.\ A {\bf 6}, 2039 (1991).
}

\lref\DKKMMS{
  M.~R.~Douglas, I.~R.~Klebanov, D.~Kutasov, J.~Maldacena,
  E.~Martinec and N.~Seiberg,
  ``A new hat for the c = 1 matrix model,''
  arXiv:hep-th/0307195.
}

\lref\GOW{
  P.~Goddard, D.~I.~Olive and G.~Waterson,
  ``Superalgebras, Symplectic Bosons And The Sugawara Construction,''
  Commun.\ Math.\ Phys.\  {\bf 112}, 591 (1987).
}

\lref\BCMN{
  P.~Bouwknegt, A.~Ceresole, J.~G.~McCarthy and P.~van Nieuwenhuizen,
  ``The Extended Sugawara Construction For The Superalgebras SU(M+1)/(N+1).
  Part 1: Free Field Representation And Bosonization Of Super Kac-Moody
  Currents,''
  Phys.\ Rev.\ D {\bf 39}, 2971 (1989);
 ``The Extended Sugawara Construction For The Superalgebras SU(M/N). Part 2.
 The Third Order Casimir Algebra,''
 Phys.\ Rev.\ D {\bf 40}, 415 (1989).
 }

\lref\NSW{
  K.~S.~Narain, M.~H.~Sarmadi and E.~Witten,
  ``A Note On Toroidal Compactification Of Heterotic String Theory,''
  Nucl.\ Phys.\ B {\bf 279}, 369 (1987).
}

\lref\PolII{
  J.~Polchinski,
  ``String theory. Vol. 2: Superstring theory and beyond,''
}

\lref\TTU{P.~Kraus and F.~Larsen, ``Boundary string field theory of
the DD-bar system,'' Phys.\ Rev.\ D {\bf 63}, 106004 (2001)
[arXiv:hep-th/0012198];
 T.~Takayanagi, S.~Terashima and T.~Uesugi,
``Brane-antibrane action from boundary string field theory,'' JHEP
{\bf 0103}, 019 (2001) [arXiv:hep-th/0012210].
}

\lref\Sugimoto{S.~Sugimoto, ``Anomaly cancellations in type I
D9-D9-bar system and the USp(32)  string theory,'' Prog.\ Theor.\
Phys.\  {\bf 102}, 685 (1999) [arXiv:hep-th/9905159].
}

\lref\Hellerman{S.~Hellerman, ``On the landscape of superstring
theory in D $>$ 10,'' arXiv:hep-th/0405041.
}

\lref\TokunagaPJ{T.~Tokunaga, ``String theories on flat
supermanifolds,'' arXiv:hep-th/0509198.
}

\lref\Witten{E.~Witten, ``Perturbative gauge theory as a string
theory in twistor space,'' Commun.\ Math.\ Phys.\  {\bf 252}, 189
(2004) [arXiv:hep-th/0312171].
}

\lref\Sethi{S.~Sethi, ``Supermanifolds, rigid manifolds and mirror
symmetry,'' Nucl.\ Phys.\ B {\bf 430}, 31 (1994)
[arXiv:hep-th/9404186].
}

\lref\ASchwarz{A.~Schwarz,``Sigma models having supermanifolds as
target spaces,'' Lett.\ Math.\ Phys.\  {\bf 38}, 91 (1996)
[arXiv:hep-th/9506070].
}

\lref\Aganagic{M.~Aganagic and C.~Vafa, ``Mirror symmetry and
supermanifolds,'' arXiv:hep-th/0403192;
A.~Neitzke and C.~Vafa, ``N = 2 strings and the twistorial
Calabi-Yau,'' arXiv:hep-th/0402128.
}

\lref\Seki{S.~Seki and K.~Sugiyama, ``Gauged linear sigma model on
supermanifold,'' arXiv:hep-th/0503074.
}

\lref\DV{R.~Dijkgraaf and C.~Vafa, ``N = 1 supersymmetry,
deconstruction, and bosonic gauge theories,'' arXiv:hep-th/0302011.
}

\lref\KKM{H.~Kawai, T.~Kuroki and T.~Morita, ``Dijkgraaf-Vafa theory
as large-N reduction,'' Nucl.\ Phys.\ B {\bf 664}, 185 (2003)
[arXiv:hep-th/0303210].
}

\lref\PW{J.~Polchinski and E.~Witten, ``Evidence for Heterotic -
Type I String Duality,'' Nucl.\ Phys.\ B {\bf 460}, 525 (1996)
[arXiv:hep-th/9510169].
}

\lref\duality{E.~Witten, ``String theory dynamics in various
dimensions,'' Nucl.\ Phys.\ B {\bf 443}, 85 (1995)
[arXiv:hep-th/9503124].
}

\lref\VF{C.~Vafa, ``Evidence for F-Theory,'' Nucl.\ Phys.\ B {\bf
469}, 403 (1996) [arXiv:hep-th/9602022].
}

\lref\Sen{A.~Sen, ``F-theory and Orientifolds,'' Nucl.\ Phys.\ B
{\bf 475}, 562 (1996) [arXiv:hep-th/9605150].
}

\lref\GS{M.~B.~Green and J.~H.~Schwarz, ``Anomaly Cancellation In
Supersymmetric D=10 Gauge Theory And Superstring Theory,'' Phys.\
Lett.\ B {\bf 149}, 117 (1984).
}

\lref\het{D.~J.~Gross, J.~A.~Harvey, E.~J.~Martinec and R.~Rohm,
``The Heterotic String,'' Phys.\ Rev.\ Lett.\  {\bf 54}, 502 (1985);
``Heterotic String Theory. 1. The Free Heterotic String,'' Nucl.\
Phys.\ B {\bf 256}, 253 (1985);
``Heterotic String Theory. 2. The Interacting Heterotic String,''
Nucl.\ Phys.\ B {\bf 267}, 75 (1986).
}

\lref\DZ{M.~R.~Gaberdiel and B.~Zwiebach, ``Exceptional groups from
open strings,'' Nucl.\ Phys.\ B {\bf 518}, 151 (1998)
[arXiv:hep-th/9709013];
O.~DeWolfe and B.~Zwiebach, ``String junctions for arbitrary Lie
algebra representations,'' Nucl.\ Phys.\ B {\bf 541}, 509 (1999)
[arXiv:hep-th/9804210].
}

\lref\GreenSP{M.~B.~Green, J.~H.~Schwarz and E.~Witten,
``Superstring Theory. Vol. 1: Introduction.''}

\lref\FrappatPB{ L.~Frappat, P.~Sorba and A.~Sciarrino, ``Dictionary
on Lie superalgebras,'' arXiv:hep-th/9607161; `Dictionary on Lie
algebras and superalgebras,'' {\it Academic Press (2000) 410 p.}
}

\lref\DeWittCY{B.~S.~DeWitt, ``Supermanifolds,'' {\it Cambridge, UK:
Univ. Pr. (1992) 407 p. (Cambridge monographs on mathematical
physics)}. }

\lref\FreundWS{ P.~G.~O.~Freund, ``Introduction To Supersymmetry,''
{\it Cambridge, Uk: Univ. Pr. ( 1986) 152 P. ( Cambridge Monographs
On Mathematical Physics)}.
}

\lref\Kac{V.~G.~Kac, ``Lie superalgebras,'' Adv.Math.{\bf
26}(1977)8; ``A sketch of Lie superalgebra theory,'' Commun.\ Math.\
Phys.\  {\bf 53}, 31 (1977).
}

\lref\Sr{M.~Srednicki, ``IIB or not IIB,'' JHEP {\bf 9808}, 005
(1998) [arXiv:hep-th/9807138].
}

\lref\Senba{A.~Sen, ``Tachyon condensation on the brane antibrane
system,'' JHEP {\bf 9808}, 012 (1998) [arXiv:hep-th/9805170].
}

\lref\SPJ{ E.~Witten, ``Toroidal compactification without vector
structure,'' JHEP {\bf 9802}, 006 (1998) [arXiv:hep-th/9712028];

M.~Bershadsky, T.~Pantev and V.~Sadov, ``F-theory with quantized
fluxes,'' Adv.\ Theor.\ Math.\ Phys.\  {\bf 3}, 727 (1999)
[arXiv:hep-th/9805056];

Y.~Imamura, ``String junctions on backgrounds with a positively
charged orientifold plane,'' JHEP {\bf 9907}, 024 (1999)
[arXiv:hep-th/9905059];

J.~Hashiba, K.~Hosomichi and S.~Terashima, ``String junctions in B
field background,'' JHEP {\bf 0009}, 008 (2000)
[arXiv:hep-th/0005164];

J.~de Boer, R.~Dijkgraaf, K.~Hori, A.~Keurentjes, J.~Morgan,
D.~R.~Morrison and S.~Sethi, ``Triples, fluxes, and strings,'' Adv.\
Theor.\ Math.\ Phys.\  {\bf 4}, 995 (2002) [arXiv:hep-th/0103170].
}

\lref\CHL{S.~Chaudhuri, G.~Hockney and J.~D.~Lykken, ``Maximally
supersymmetric string theories in D $<$ 10,'' Phys.\ Rev.\ Lett.\
{\bf 75}, 2264 (1995) [arXiv:hep-th/9505054];
S.~Chaudhuri and J.~Polchinski, ``Moduli space of CHL strings,''
Phys.\ Rev.\ D {\bf 52}, 7168 (1995) [arXiv:hep-th/9506048].
}

\lref\PolR{J.~Polchinski,``What is string theory?,''
arXiv:hep-th/9411028.
}

\lref\MV{ J.~McGreevy and H.~Verlinde, ``Strings from tachyons: The
c = 1 matrix reloaded,'' JHEP {\bf 0312}, 054 (2003)
[arXiv:hep-th/0304224];
I.~R.~Klebanov, J.~Maldacena and N.~Seiberg, ``D-brane decay in
two-dimensional string theory,'' JHEP {\bf 0307}, 045 (2003)
[arXiv:hep-th/0305159];
J.~McGreevy, J.~Teschner and H.~L.~Verlinde, ``Classical and quantum
D-branes in 2D string theory,'' JHEP {\bf 0401}, 039 (2004)
[arXiv:hep-th/0305194];
A.~Sen, ``Open-closed duality: Lessons from matrix model,'' Mod.\
Phys.\ Lett.\ A {\bf 19}, 841 (2004) [arXiv:hep-th/0308068].
T.~Takayanagi and S.~Terashima, ``c = 1 matrix model from string
field theory,'' arXiv:hep-th/0503184.
}

\lref\TT{T.~Takayanagi and N.~Toumbas, ``A matrix model dual of type
0B string theory in two dimensions,'' JHEP {\bf 0307}, 064 (2003)
[arXiv:hep-th/0307083]
}

\lref\GTT{
 S.~Gukov, T.~Takayanagi and N.~Toumbas, ``Flux backgrounds in 2D
string theory,'' JHEP {\bf 0403}, 017 (2004) [arXiv:hep-th/0312208].
}

\lref\PolchinskiMT{
  J.~Polchinski,
  ``Dirichlet-Branes and Ramond-Ramond Charges,''
  Phys.\ Rev.\ Lett.\  {\bf 75}, 4724 (1995)
  [arXiv:hep-th/9510017].
}

\lref\MaldacenaRE{
  J.~M.~Maldacena,
  ``The large N limit of superconformal field theories and supergravity,''
  Adv.\ Theor.\ Math.\ Phys.\  {\bf 2}, 231 (1998)
  [Int.\ J.\ Theor.\ Phys.\  {\bf 38}, 1113 (1999)]
  [arXiv:hep-th/9711200].
}

\lref\AganagicDB{
  M.~Aganagic, A.~Klemm, M.~Marino and C.~Vafa,
  ``The topological vertex,''
  Commun.\ Math.\ Phys.\  {\bf 254}, 425 (2005)
  [arXiv:hep-th/0305132].
}

\lref\BerkovitsIM{
  N.~Berkovits, C.~Vafa and E.~Witten,
``Conformal field theory of AdS background with Ramond-Ramond
flux,''
  JHEP {\bf 9903}, 018 (1999)
  [arXiv:hep-th/9902098].
}

\lref\Ber{M.~Bershadsky, S.~Zhukov and A.~Vaintrob, ``PSL(n$|$n)
sigma model as a conformal field theory,''
 Nucl.\ Phys.\ B {\bf 559}, 205 (1999)
  [arXiv:hep-th/9902180].
}

\lref\HoWi{
  P.~Horava and E.~Witten,
  ``Heterotic and type I string dynamics from eleven dimensions,''
  Nucl.\ Phys.\ B {\bf 460}, 506 (1996)
  [arXiv:hep-th/9510209];
``Eleven-Dimensional Supergravity on a Manifold with Boundary,''
  Nucl.\ Phys.\ B {\bf 475}, 94 (1996)
  [arXiv:hep-th/9603142].
}

\lref\Yost{
  S.~A.~Yost,
  ``Supermatrix models,''
  Int.\ J.\ Mod.\ Phys.\ A {\bf 7}, 6105 (1992)
  [arXiv:hep-th/9111033].
}

\lref\HS{J.~A.~Harvey and A.~Strominger,
  ``The heterotic string is a soliton,''
  Nucl.\ Phys.\ B {\bf 449}, 535 (1995)
  [Erratum-ibid.\ B {\bf 458}, 456 (1996)]
  [arXiv:hep-th/9504047].
}

\lref\Senhet{
  A.~Sen,
  ``String string duality conjecture in six-dimensions and charged solitonic
  strings,''
  Nucl.\ Phys.\ B {\bf 450}, 103 (1995)
  [arXiv:hep-th/9504027].
}

\lref\FS{L.~Frappat and A. Sciarrino,
  ``Hyperbolic Kac-Moody superalgebras,''
  [arXiv:math-ph/0409041].
}

\lref\GC{N.~Arkani-Hamed, H.~C.~Cheng, M.~A.~Luty and S.~Mukohyama,
  ``Ghost condensation and a consistent infrared modification of gravity,''
JHEP {\bf 0405}, 074 (2004) [arXiv:hep-th/0312099];
N.~Arkani-Hamed, P.~Creminelli, S.~Mukohyama and M.~Zaldarriaga,
  ``Ghost inflation,''
  JCAP {\bf 0404}, 001 (2004)
  [arXiv:hep-th/0312100].
}

\lref\WittenIM{
  E.~Witten,
  ``Bound states of strings and p-branes,''
  Nucl.\ Phys.\ B {\bf 460}, 335 (1996)
  [arXiv:hep-th/9510135].
}

\lref\PolchinskiBG{
  J.~Polchinski,
  ``Open heterotic strings,''
  arXiv:hep-th/0510033.
}

\lref\GIR{
  D.~Gaiotto, N.~Itzhaki and L.~Rastelli,
  ``On the BCFT description of holes in the c = 1 matrix model,''
  Phys.\ Lett.\ B {\bf 575}, 111 (2003)
  [arXiv:hep-th/0307221].
}

\lref\OOY{
  H.~Ooguri, Y.~Oz and Z.~Yin,
  ``D-branes on Calabi-Yau spaces and their mirrors,''
  Nucl.\ Phys.\ B {\bf 477}, 407 (1996)
  [arXiv:hep-th/9606112].
}

\lref\BSS{
  T.~Banks, N.~Seiberg and E.~Silverstein,
  ``Zero and one-dimensional probes with N = 8 supersymmetry,''
  Phys.\ Lett.\ B {\bf 401}, 30 (1997)
  [arXiv:hep-th/9703052].
}

\lref\Morris{T.~R.~Morris, ``A manifestly gauge invariant exact
renormalization group,'' arXiv:hep-th/9810104;
  ``A gauge invariant exact renormalization group. I and II''
  Nucl.\ Phys.\ B {\bf 573}, 97 (2000)
  [arXiv:hep-th/9910058],
JHEP {\bf 0012}, 012 (2000)
  [arXiv:hep-th/0006064].
}

\lref\EMR{N.~Evans, T.~R.~Morris and O.~J.~Rosten, ``Gauge Invariant
Regularization in the AdS/CFT Correspondence and Ghost D-branes,''
arXiv:hep-th/0601114.
}

\lref\Caldwell{R.~R.~Caldwell, ``A Phantom Menace?,'' Phys.\ Lett.\
B {\bf 545}, 23 (2002) [arXiv:astro-ph/9908168].
}

\lref\Gibbons{G.~W.~Gibbons and D.~A.~Rasheed, ``Dyson Pairs and
Zero-Mass Black Holes,'' Nucl.\ Phys.\ B {\bf 476}, 515 (1996)
[arXiv:hep-th/9604177];
 G.~W.~Gibbons,
  ``Phantom matter and the cosmological constant,''
  arXiv:hep-th/0302199.
}

\lref\KaSu{D.~E.~Kaplan and R.~Sundrum, ``A symmetry for the
cosmological constant,'' arXiv:hep-th/0505265.
}

\baselineskip 18pt plus 2pt minus 2pt

\Title{\vbox{\baselineskip12pt \hbox{hep-th/0601024} \hbox{NSF-KITP-05-101}
  }}
{\vbox{\centerline{Ghost D-branes}}}

\centerline{Takuya Okuda\foot{e-mail: takuya at kitp.ucsb.edu} and
Tadashi Takayanagi\foot{e-mail: takayana at kitp.ucsb.edu}}

\medskip\centerline{Kavli Institute for Theoretical
Physics}\centerline{University of California} \centerline{Santa
Barbara, CA 93106 USA}

\vskip .5in \centerline{\bf Abstract}

We define a ghost D-brane in superstring theories as an object that
cancels the effects of an ordinary D-brane. The supergroups $U(N|M)$
and $OSp(N|M)$ arise as gauge symmetries in the supersymmetric
world-volume theory of D-branes and ghost D-branes. A system with a
pair of D-brane and ghost D-brane located at the same location is
physically equivalent to the closed string vacuum. When they are
separated, the system becomes a new brane configuration. We
generalize the type I/heterotic duality by including $n$ ghost
D9-branes on the type I side and by considering the heterotic string
whose gauge group is $OSp(32+2n|2n)$. Motivated by the type IIB
S-duality applied to D9- and ghost D9-branes, we also find type II-like
closed superstrings with $U(n|n)$ gauge symmetry.

\noblackbox

\Date{January, 2006}

\listtoc
\writetoc

\newsec{Introduction and Summary}

D-branes \PolchinskiMT\ have been the basis of most developments in
string theory for the last ten years. The AdS/CFT correspondence
\MaldacenaRE, for example, was discovered by considering the
near-horizon limit of D3-branes in type IIB string theory. Many of
exact calculations in topological string theory have also been made
possible by the inclusion of D-branes \AganagicDB. The
identification of  fermions with D0-branes revivided the
investigation of two dimensional string theories \MV \TT \DKKMMS.

In this paper, we introduce the notion of ghost D-branes in
superstring theories. We define a ghost D-brane as an object that
cancels the effects of an ordinary D-brane. We call them ghost
D-branes because the gauge fields and transverse scalars on them
have the wrong signs in their kinetic terms. The open strings
between a D-brane and a ghost D-brane have the opposite statistics
relative to the usual ones. Ghost D-branes replace ordinary
Chan-Paton matrices with supermatrices. The Lie supergroups $U(N|M)$
and $OSp(N|M)$ arise as gauge groups in this way. In particular,
this leads to a new type I string theory with gauge group
$OSp(32+2n|2n)$ and type IIB string with gauge group $U(n|n)$ with
any positive integer $n$. A similar idea can clearly be applied to
define ghost M2 and M5-branes in M-theory.

We stress that ghost D-branes are different from anti D-branes. The
boundary state of a ghost D-brane is minus the whole boundary state
of an ordinary D-brane, whereas an anti D-brane has a minus sign
just in the RR sector. Thus ghost  D-branes have negative tension
and anti-gravitate. A ghost D-brane completely cancels the effects
of a D-brane when they are on top of each other with a trivial gauge
field background. When $M$ ghost D-branes are coincident with
$N(\geq M)$ D-branes, the system is equivalent to the one with $N-M$
D-branes with no ghost branes as we show explicitly. More subtle are
the situations where we separate ghost D-branes from D-branes or
turn on non-trivial gauge backgrounds like Wilson lines. The
supergravity solution for isolated ghost D-branes appears to be
singular near the branes essentially because they are
anti-gravitating sources. A system with ghost branes have two kinds
of ghost-like fields: some of them have the wrong spin-statistics
relations while the others have the wrong signs in their kinetic
terms. The theory will not be unitary when ghost branes are not
canceled by ordinary branes. Once the fields on 
ghost branes are excited, there may be instabilities because of
the wrong signs in the kinetic terms. We expect that
despite these pathologies the concept of ghost branes
and their cancellation against ordinary branes will be useful tools.

Ghost D-branes preserve the same  supercharges as ordinary D-branes,
and it is natural to consider its strong coupling dual. Indeed we
will show that the  type I $OSp(32+2n|2n)$ string theory is dual to
a heterotic  $OSp(32+2n|2n)$ string theory (see \TokunagaPJ\ for an
earlier discussion on this heterotic string) by generalizing the
 type I/heterotic duality \PW.
We will also discuss a supergroup extension of the $E_8\times E_8$
heterotic string. This heterotic string has infinitely many massless
gauge fields. We expect that it is uplifted to the Horava-Witten
\HoWi\ like setup in M-theory. It is also interesting to consider
the type IIB S-duality applied to $n$ pairs of D9- and ghost
D9-branes. As its strong coupling limit, we find a novel
superstrings that have the $U(n|n)$ gauge fields in the closed
string sector. This superstring theory looks like a heterotic
modification of the ordinary type IIB string, which is equivalent to
the type IIB string with $n$ NS9-branes and $n$ ghost NS9-branes.

Another context where various Lie groups appear is the study of
string junctions suspended between $(p,q)$ 7-branes (refer to
e.g.\DZ\ and references therein). It is natural to add ghost
7-branes and ask which Lie supergroups can appear as symmetry groups. We
will realize $SU(N|M)$ and $OSp(2M|2N)$ symmetries in this way and
obtain their Dynkin diagrams in terms of string junctions.

There is a rather similar example in two dimensional
 string theory\foot{A closely related remark was also made in
 \DKKMMS, where the boundary state for a hole state was
 considered.}. The $c=1$ matrix model dual provides its non-perturbative
definition. This model is equivalent to the quantum mechanics of
infinitely many free fermions in an inverse harmonic potential. The
ordinary vacuum of two dimensional string theory corresponds to the
Fermi surface $\f{1}{2}(p^2-x^2)=E_0$. The fluctuations on this
fermi surface correspond to the
 massless scalar field
in the closed string theory (e.g., see the review \PolR\ and
references therein).

If we get rid of a band of fermions with the energy $E_2\leq E\leq
E_1$ assuming $E_1<E_0$, then we have
 three Fermi surfaces at
$E=E_0,\  E=E_1$ and $E=E_2$. Thus this system possesses three
massless bosons $\vp_0,\vp_1$ and $\vp_2$ as fluctuations (or
collective fields) on these Fermi surfaces (see Fig.1).
Interestingly, the second one $\vp_1$ turns out to be a ghost, i.e.
it has the wrong sign in  its kinetic term.  This is because the
higher energy region is completely filled at the Fermi surface
$E=E_1$ and excitations always have negative energy. On the other
hand, since the opposite is true at $E=E_{0}$ and $E=E_{2}$, the
fields $\vp_0$ and $\vp_2$ are ordinary massless scalar fields.

Despite its seeming instability near $E=E_1$, this system is stable
because the fermions in the $c=1$ matrix model are free. To be
exact, we need to consider type 0 string theory \TT \DKKMMS\ to
obtain its non-perturbative completion. In the limit $E_1\to E_2$
where the band disappears, the scalar field $\vp_1$ and the ghost
field $\vp_2$ cancel out as expected. In this way we find a
physically sensible theory with ghosts. Then we can apply the modern
identification of the fermions with D0-branes in two dimensional
string theory \MV \DKKMMS\ \GIR\ to find what background in two
dimensional string this configuration corresponds to. Removing
fermions with energy $E_2\leq E\leq E_1$ can be interpreted as
condensing infinitely many ghost D0-branes. Notice also that we
cannot describe such a background in the two dimensional effective
dilaton-gravity theory\foot{A similar situation appears in the flux
background of the two dimensional type 0B string theory. A
description of this background was proposed in \GTT\ using a quantum
field redefinition of the RR scalar field.}.

\fig{In the $c=1$ matrix model, a ghost field can appear as a
collective field when we remove a band of
fermions.}{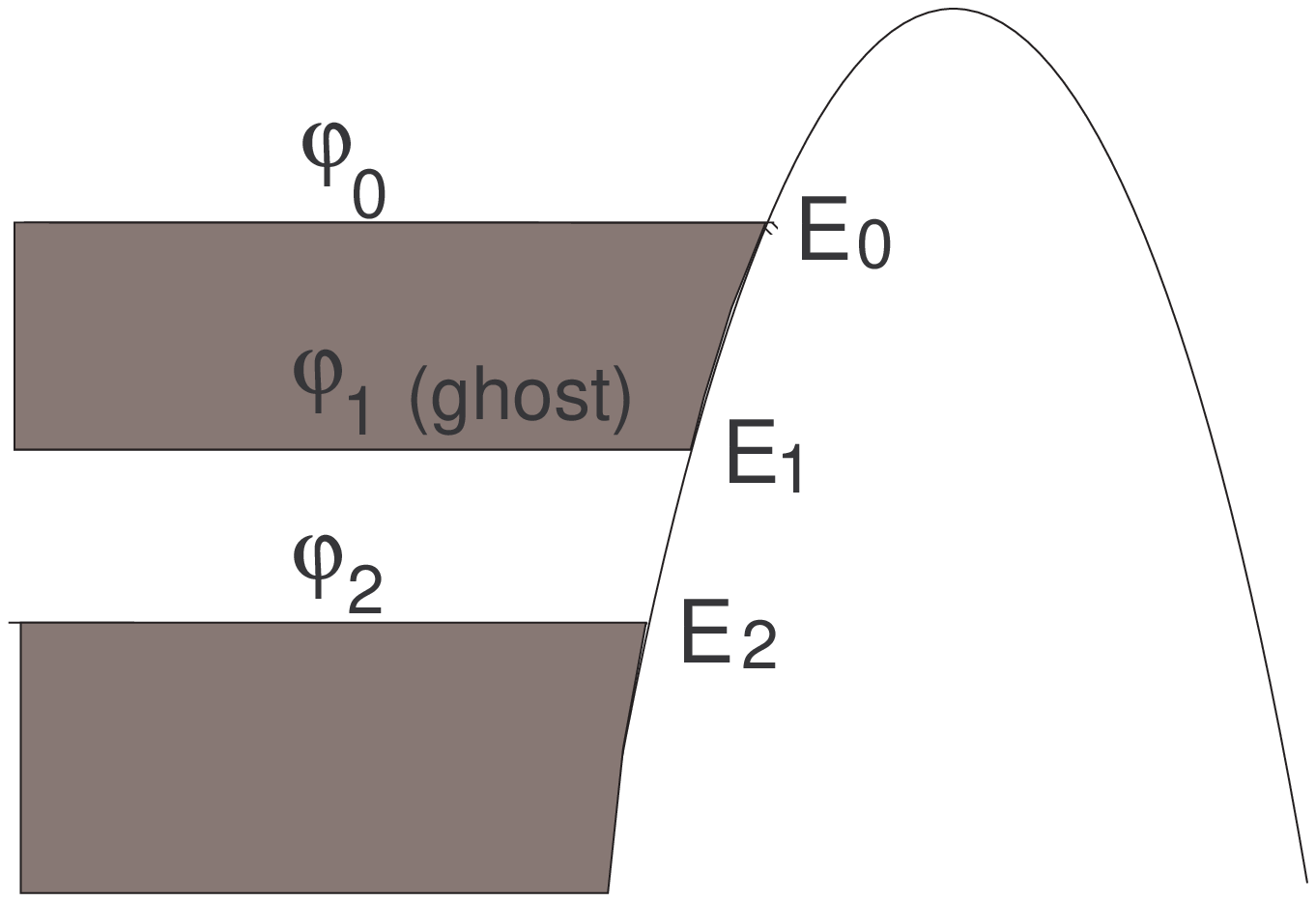}{2.5truein}

Also closely related to ghost D-branes are the anti D-branes in
topological string theory \V. Indeed the amplitudes of the anti
D-branes in topological string theory precisely cancel the
amplitudes of the D-branes. The cancellation between topological
branes and anti branes has been heavily used in the recent
developments in topological string theory \AganagicDB. It seems that
anti D-branes in topological string theory are more directly a
counterpart of ghost D-branes than anti D-branes in physical
superstring theory.

The paper is organized as follows. In section two, we demonstrate
that the Lie superalgebras $U(N|M)$ and $OSp(N|M)$ arise in open
string theory by including ghost D-branes. In section three, we
discuss this open string theory and show that the branes and ghost
branes cancel each other. We do this by demonstrating that theories
with supergroup symmetries reduce to theories with ordinary bosonic
symmetries. This part does not require the knowledge of string
theory and should be readable by anyone who knows quantum field
theory. Also we prove that the gauge anomaly is canceled in the type
I theory. In section 4 we study the strong coupling limit of the
type I $OSp(32+2n|2n)$ string theory and argue that it is given by
the heterotic $OSp(32+2n|2n)$ string theory. We also discuss a
supergroup extension of the heterotic $E_8\times E_8$ string theory
and mention its M-theory origin. Section 5 is devoted to the
construction of type IIB-like superstrings with $U(N|N)$ gauge
symmetry, motivated by S-duality in type IIB string theory. In
section 6 we `derive' Dynkin diagrams of Lie superalgebras from
7-branes configurations in type IIB string theory. In section 7 we
discuss other implications of our results and future directions.

\newsec{Chan-Paton Factors and Lie Superalgebras}

Usually, the gauge group which appears in the open string sector
is $U(N)$ in type II string theory, and $SO(N)$ or $Sp(N)$ in type I
string theory. These are realized by assigning Chan-Paton matrices to
open strings. We will generalize these Chan-Paton
matrices in superstring theories into elements of Lie superalgebras.
For reviews of Lie superalgebras refer to \FrappatPB \DeWittCY
\FreundWS\ as well as appendix A of the present paper.
 Our results in this section, section 3 and section 6
 can also be applied to
bosonic and type 0 strings. Though in this paper we only consider
BPS configurations of D-branes, it is straightforward to apply
similar arguments to non-BPS configurations such as
brane-antibrane systems \Senba.

\subsec{$U(N|M)$ and Type II String Theory}

We argue that this extension corresponds the inclusion of negative
tension D$p$-branes. We call them ghost D$p$-brane. In the boundary
state formalism, an ordinary BPS D-brane is represented by a
boundary state $|D\lb=|D\lb_{\rm NSNS}+|D\lb_{\rm RR}$. The ghost
D$p$-brane is defined simply by the boundary state $|gD\lb$ with an
overall minus sign \eqn\BPSD{|gD\lb=-|D\lb.} It is crucial to
distinguish a ghost D-brane from an anti D-brane. The boundary state
for the latter is obtained by flipping the sign of the RR part only
\Senba. A brane anti-brane system is non-supersymmetric, while a
brane ghost-brane system possesses sixteen supersymmetries. This is
because a ghost D-brane keeps the same half supersymmetries as the
BPS D-brane does.

Consider an cylinder amplitude between a D$p$-brane and a ghost
D$p$-brane in type II string theory. This is obviously given by
minus the ordinary amplitude between BPS D$p$-branes \eqn\amp{\la
gD|\Delta|D\lb =-\la D|\Delta|D\lb,} where $\Delta$ is the closed
string propagator. Interpreted in the open string channel, the
spectrum is given by replacing bosons in the ordinary spectrum on
D-branes with fermions and vice versa. Thus all fields between a
D$p$-brane and a ghost D$p$-brane are ghost-like. The gluons become
fermions and the gauginos become bosons. Now consider $N$
D$p$-branes and $M$ ghost D$p$-branes on top of each other. We can
summarize this field content by a hermitian supermatrix (see
appendix A) \eqn\supermatga{\Phi = \left(\matrix{\phi_1&\psi\cr
\psi^\dagger &\phi_2}\right),} where $\phi_1$ and $\phi_2$ are
bosonic hermitian matrices, while $\psi$ is a complex fermionic
matrix. The diagonal part $\phi_1$ (or $\phi_2$) corresponds to the
open strings between D-branes (or ghost D-branes). The off-diagonal
part $\psi$ corresponds to the open strings between D-branes and
ghost D-branes and thus they have the opposite statistics. In other
words, the world-volume theory on $N$ D$p$-branes and $M$ ghost
D$p$-branes is given in low energy by a $p+1$ dimensional super
Yang-Mills theory with gauge group $U(N|M)$. In this way, a $U(N|M)$
valued Chan-Paton factor naturally appears. If we place $N$
D9-branes and $N$ ghost 9-branes, the tadpoles are canceled and we
obtain a consistent ten dimensional $U(N|N)$ super Yang-Mills theory
coupled to type IIB supergravity\foot{In the presence of $N$
D9-branes and $N$ anti D9-branes, the system is described by the
$U(N)\times U(N)$ gauge theory with bi-fundamental (complex) tachyon
fields \Senba \Sr. Clearly there is no $U(N|N)$ symmetry. Instead,
the mathematical structure of the gauge theory is described by the
Quillen's superconnection, where odd elements correspond to tachyon
fields (and not the gauge field) as shown in \TTU.}. Note that in
this case the NSNS tadpole is also zero and therefore we do not need
to invoke the Fischler-Susskind mechanism.

The appearance of the supergroup $U(N|M)$ in a similar way was
mentioned in \DKKMMS\ in the context of matrix model duals of two
dimensional string theories. In the topological string context, the
gauge group of the brane-anti brane system was argued to be $U(N|M)$
in \V. This is consistent with our discussion because in topological
string theory we only have the RR-sector part \OOY\ and thus the
boundary state for an anti D-brane is precisely minus that of a
D-brane.

\subsec{$OSp(N|M)$ and Type I String Theory}

To define type I string theory, we need an orientation projection.
D1 or D9-branes give an $SO(N)$ Chan-Paton factor, while
D5-branes\foot{Our convention for symplectic groups is such that
$Sp(2)=SU(2)$. Sometimes $Sp(N)$ is denoted by $USp(N)$.} give
$Sp(N)$ ($N$ is even) \PolII. Since there exists an orientifold
9-plane in the background, the brane configuration is described by
the sum \eqn\typeone{|D\lb+|\Omega\lb,} where $|\Omega\lb$ is the
crosscap state. From the overlap of two of such states we get the
cylinder + M\"obius strip + Klein bottle amplitudes. We extend this
theory by including negative tension D$p$-branes as before. Notice
that the $|\Omega\lb$ part remains the same because we do not want
to modify the theory itself. Then the ghost brane configuration is
described by \eqn\typeoness{-|D\lb+|\Omega\lb.}

Consider the system of $N$ D9 (or D1)-branes and $M/2$ ghost D9 (or
D1)-branes, assuming $M$ is even\foot{The reason why we have to mod
the number of ghost 9-branes by the factor 2 is that the $Sp$
projection will be imposed on them as we will see shortly and
therefore they only make sense when $M$ is an even integer.}.
For open strings between two ghost D9 branes, then it is
obvious from \typeoness\ that we get an extra minus sign for the
M\"obius amplitudes. This means that the Chan-Paton factor for the
ghost D9 branes is in $Sp(M)$. The open strings between D9 and ghost D9
become ghost-like as before. We thus find the $OSp(N|M)$
gauge group in type I string theory. On the other hand,  $N$
D5-branes and $M$ ghost D5-branes give rise to the gauge group
$OSp(M|N)$ with an opposite sign for the coupling constant
$g_{YM}^2$. The detailed structure of $OSp(N|M)$ will be discussed
in the next subsection and in appendix A.

For 9-branes,
we need to impose the condition $N-M=32$ so that the tadpoles are canceled.
As we will see later,
indeed the gauge anomaly  is canceled in super Yang-Mills theory
with  gauge group $OSp(32+2n|2n)$ by the Green-Schwarz
mechanism  \GS\foot{With $32+n$ D9-branes and $n$ anti
D9-branes, the gauge group becomes $SO(32+n)\times SO(n)$ (or
$Sp(32+n)\times Sp(n)$ by considering the opposite $\Omega$
projection) \Sugimoto \Hellerman.}.

\subsec{$\Omega$ Action and Lie Superalgebras}

So far we have been discussing ghost D-branes from the
viewpoint of boundary states. Here
we consider how the orientation projection $\Omega$
acts on open string states.

The $\Omega$ action reverses the orientation of the world-sheet as
usual. It always acts on the oscillator part of a massless gluon
state as $\Omega~|{\rm gluon}\lb=-|{\rm gluon}\lb.$ Below we
concentrate on its action on the Chan-Paton factor. Since it
exchanges the two boundaries of the open string world-sheet,
$\Omega$ will act like $|i,j\lb \to |j,i\lb$, where
$i,j=1,2,\ddd,N+M$ in the presence of $N$ D-branes and $M/2$
ghost-branes. We express the Chan-Paton matrix by $\lambda$ and
assume that $\lambda$ is hermitian as usual. We expect that $\Omega$
acts by transposing $\lambda$. When we $\lambda$ is a supermatrix we
need to modify the definition of a transposed matrix. Indeed the
effective Yang-Mills action on branes is $S\propto {\rm
Str}F^2+\ddd$ and the $\Omega$ action should be a symmetry of this
system. Because in general $(\lambda_1\lambda_2)^{T}\neq
\lambda_2^T\lambda_1^T$ for supermatrices, $\lambda\to \lambda^T$ is
not a symmetry.
 As is explained  in  appendix A,
we should supertranspose $\lambda$ to $\lambda^{\T}$. This satisfies
$(\lambda_1\lambda_2)^{{\ti T}}= \lambda_2^{\ti T}\lambda_1^{\ti T}$
and $\Str \lambda=\Str \lambda^\T$. Notice also that $\ti{T}$ is not
a $\IZ_2$ action since $(\lambda^{\ti T})^{\ti T}=K\lambda K$, where
$K={\rm diag}(I_{N},-I_{M})$. The matrix $I_{M}$ denotes the
$M\times M$ identity matrix.

Then we can write the $\Omega$ action in the following form
\eqn\Omegaction{\Omega: \lambda\to
\gamma^{-1}\lambda^{\ti{T}}\gamma,} where $\gamma$ is a $U(N|M)$
matrix corresponding to a gauge transformation on D-branes. By requiring
that this is a $\IZ_2$ action $\Omega^2=1$, we find the
condition\foot{ Here we employed Schur's lemma for
supermatrices \FrappatPB. Also we assumed the generic situation $N\neq M$.}
 for $\gamma$
\eqn\constr{\gamma^{\ti{T}}=\pm \gamma~ K.} Under the $U\in U(N|M)$
rotation $\lambda\to U\lambda U^{-1}$, the matrix $\gamma$
transforms as $\gamma\to U\gamma U^{\ti{T}}$. We can pick up the
following solution to the constraint \constr\
\eqn\sol{\eqalign{\gamma_{OSp}&= \left(\matrix{I_{N}&0\cr
0&\eta_M}\right), \cr \gamma_{SpO}&= \left(\matrix{\eta_{N}&0\cr
0&I_M}\right) ,}} corresponding to the $\pm$ sign in \constr,
respectively. The $M\times M$ matrix $\eta_M$ is defined by
$\eta=-\sigma_2\otimes I_{M/2}$ (see also (A.14) in appendix A).

The massless gauge field should satisfy $\Omega=-1$ under the action
\Omegaction.
In the case $\gamma=\gamma_{OSp}$, this leads precisely to the condition that the
Chan-Paton factor is a generator of $OSp(N|M)$:
\eqn\condfina{\lambda=-\gamma_{OSp}~\lambda^{\ti{T}}~\gamma_{OSp}^{-1}.}
This is realized when we consider ghost D9 or D1-branes in type
I string theory\foot{In the ghost D1-branes case,
the transverse scalars are not in the adjoint representation
 and are subject to the the opposite projection $\Omega=1$, i.e.,
they satisfy $\lambda=-\gamma_{SpO}~\lambda^{\ti{T}}~\gamma_{SpO}^{-1}$.}. The
other case $\gamma_{SpO}$ corresponds to $OSp(M|N)$ and
occurs when we consider D5-branes.

\newsec{World-Volume Theory on Branes and Ghost Branes}

The world-volume theory on $N$ D$p$-branes and $M$ (or $M/2$) ghost
D$p$-branes can clearly be described by a gauge theory with $U(N|M)$
(or $OSp(N|M)$) Chan-Paton matrices. The low energy action\foot{The
DBI action takes the form $S=T_{{\rm D}p}\int d^{p+1}x~\Str
\s{-\det(1+2\pi\al F_{\mu\nu})}.$ It has a minus sign in front of
the whole DBI action
 for the gauge fields with wrong-sign kinetic terms.} is that of the
$(p+1)$ dimensional super Yang-Mills theory whose gauge group is one
of these supergroups \eqn\yangmills{S=\f{1}{g^2_{YM}}\int
d^{p+1}x~\str\left[-{1\over 4} (F_{\mu\nu})^2 -{1\over
2}(D_\mu\phi^i)^2+\ddd\right],} where $F_{\mu\nu}=\de_\mu A_{\nu}
-\de_\nu A_{\mu}-i[A_{\mu},A_\nu]$ is the field-strength and
$\phi^i$ are the transverse scalars. This action \yangmills\ is
invariant under the gauge transformation\foot{Notice that only
commutators appear in these expressions for the Yang-Mills theory.
Anti-commutators which are typical in Lie superalgebras (see
appendix A) arise when we expand a supermatrix-valued field
$\phi(x)$ by bosonic generators $T^A$ with coefficients $\phi_A(x)$.
Only $\phi_A$ can be Grassmann odd and in that case we find
anti-commutators $T^AT^B+T^B T^A$.} $\delta
A_{\mu}=\de_\mu\chi+i[\chi,A_{\mu}]$. Since the gauge field and the
transverse scalar fields take supermatrix values, some of the
bosonic modes have an extra minus sign in the kinetic terms\foot{
Refer to \Gibbons\ for classical solutions of the Einstein-Maxwell
theory with a minus sign in front of kinetic terms of gauge fields
as appear in ghost D-branes.}. Also there exists a fermionic gauge
and scalar fields. These super Yang-Mills theories possess sixteen
supersymmetries as in the ordinary case. Because of this, we can
indeed extend the various string dualities consistently as we will
explain later.

\subsec{Cancellation between Branes and Ghost Branes}

Let us assume all branes and ghost branes are situated at the same
location. In the boundary state formalism in section 2, it is clear
that the branes and ghost branes  cancel each other. Thus
 the open string theory on $N$ D$p$-branes and $M$ ghost D$p$-branes is
equivalent to the theory with $N-M$ D$p$-branes when $N\geq M$ or
the one with $M-N$ ghost D$p$-branes when $M>N$. For example, the
partition function $Z$ of the open string theory should satisfy the
following equality \eqn\formula{Z\left[U(N|M),g_{YM}^2\right]
=Z\left[U(N-M),g_{YM}^2\right],} in type II string theory and
\eqn\formulaosp{
Z\left[OSp(N|M),g_{YM}^2\right]=Z\left[O(N-M),g_{YM}^2)\right],} in
type I string theory\foot{Here we consider $p=1,9$ branes. For $p=5$
branes we have only to replace $g_{YM}^2$ with $-g_{YM}^2$ in the
right-hand side of \formulaosp.}, assuming $N\geq M$. When $N< M$ we
clearly obtain \eqn\obtains{\eqalign{Z\left[U(N|M),g_{YM}^2\right]&
=Z\left[U(M-N),-g_{YM}^2\right],\cr
Z\left[OSp(N|M),g_{YM}^2\right]&=Z\left[Sp(M-N),-g_{YM}^2\right].}}
We can also derive a similar relation for the correlation functions
of gauge invariant operators.

We can also imagine a situation with branes of different dimensionalities.
For example, consider the system of several D3-branes coincident with $N$
D7-branes and $M$ ghost D7-branes.
The world-volume theory of 3-branes now has a flavor symmetry $U(N|M)$
with $N$ usual fermionic quarks and $M$ bosonic quarks.
From the brane picture it is clear that this system is equivalent to
a theory with $N-M$ usual quarks.

These statements are non-trivial in open string field theory or
super Yang-Mills theory \yangmills.
Even though the open string theory is complicated
because of infinitely many fields and interactions, the essential
point of this cancellation  is
the combinatorics of Chan-Paton matrices. In other words, it is enough to show the
reduction for corresponding zero dimensional supermatrix models.
This is because all the other parts, e.g., propagators, integrations
of momenta and interactions, are exactly the same due to the
supergroup gauge symmetry dictates that the action is constructed out of
supertraces.
It may appear that we need to consider multi-supermatrix
models since we have many open string fields. However, as is shown in
appendix B, cancellation takes place for each Wick
contraction of correlation functions. Then it is obvious that
the reduction holds for multi-matrix models if we prove it
for one-matrix models.

This cancellation \formula\ for $U(N|M)$ matrix model has been
already shown\foot{Similar matrix models have been discussed in \DV
\KKM\ in order to compute superpotentials of 4D $N=1$ super
Yang-Mills theories.} perturbatively by Feynman diagrams \AM \Yost\
and non-perturbatively by Virasoro constraints \AM. In appendix B we
review these proofs and also extend it to the $OSp(N|M)$ case to
show \formulaosp\ and \obtains. The similar reduction of $U(N|M)$
was also explained for the supergroup sigma model in \BerkovitsIM
\Ber\ and for the topological branes \V.
In the appendix, we also show that the reduction $U(N|M)\rightarrow U(N-M)$ holds
for flavor symmetry, too.

Now, it is possible to give large vevs to $\phi^i$s so that the
D-branes are far away from the ghost D-branes. For such a
configuration the cancellations like \formula \formulaosp\ are not
true any more. In this paper we will not get into details of the
physical interpretations of these unusual modes, but only discuss
briefly in the final section.

\subsec{Anomaly Cancellation in Type I $OSp(32+2n|2n)$ String}

In order to obtain a physically sensible gauge theory, we have to
require all gauge anomalies to vanish. The ten dimensional $N=1$
super Yang-Mills theory always suffers from the hexagon anomaly.
However, when we couple the gauge theory with $N=1$ supergravity,
the anomaly is canceled by the Green-Schwarz mechanism \GS.

In this cancellation mechanism \GS \GreenSP, the essential identity
required was \eqn\anomalyone{
\Tr_{ad}[F^6]=\f{1}{48}\Tr_{ad}[F^4]\cdot\Tr_{ad}[F^2]
-\f{1}{14400}\left(\Tr_{ad}[F^2]\right)^3,} where $F$ is the gauge
field strength. In the end, we find that \anomalyone\ is satisfied
 for the celebrated gauge groups $SO(32)$ and $E_8\times E_8$.

In our case of the supergroups, we need to replace $\Tr_{ad}$ with
the supertrace $\Str_{ad}$. To see that \anomalyone\  is satisfied
in this case, we need to rewrite the supertrace $\Str_{ad}$ in the
adjoint representation in terms of the supertrace $\Str$ in the
fundamental representation.  This can be found from the explicit
form of the generators in the adjoint representation in terms of the
generators $t^A_{\mu\nu}$ in the fundamental representation:
\eqn\genera{\eqalign{(t^{A})^{\mu\nu}_{\rho\lambda} &=\f{1}{2}
\Bigl[\delta_{\lambda\nu}t^A_{\mu\rho}
-(-1)^{(|\mu|+|\lambda|)(|\lambda|+|\nu|)}
\delta_{\mu\rho}t^A_{\lambda\nu}\cr &\ \
+(-1)^{(|\mu|+|\rho|)(|\nu|+1)}
\gamma_{\mu\lambda}\gamma_{\rho\ap}t^A_{\ap\nu}
-(-1)^{|\lambda|(|\lambda|+|\nu|)}
\gamma_{\ap\lambda}\gamma_{\rho\nu}t^A_{\mu\ap}\Bigl].}} The
supertrace is defined as $\Str M=(-1)^{|\mu|}M_{\mu\mu}$ in the
fundamental representation and as
$Str_{ad}M=(-1)^{|\mu|+|\nu|}M^{\mu\nu}_{\mu\nu}$ in the adjoint
representation. Using \genera, we can see that the results for the
supergroup $OSp(N|M)$ are obtained from those for $O(N-M)$ by
replacing\foot{The same is true for $SU(N|M)$ and $SU(N-M)$.} traces
with supertraces
 (for some details see appendix B.4).
 Explicitly, we
find \eqn\anomalytwo{\eqalign{\Str_{ad}[F^2]& =(N-M-2)~\Str[F^2],
\cr \Str_{ad}[F^4]&=(N-M-8)~\Str[F^2]+3\left(\Str[F^2]\right)^2 ,\cr
\Str_{ad}[F^6]&=(N-M-32)~\Str[F^6]+15\Str[F^2]\Str[F^4].}} The
condition \anomalyone\ is satisfied only when $N-M=32$. We conclude
that the anomaly is canceled for the gauge group\foot{It will also
be interesting to check the anomaly cancellation for
$E(8+{n\over 2},{n\over 2})\times E(8+{n\over 2},{n\over 2})$ that
will be discussed in the next section.}
 $OSp(32+2n|2n)$ as we expected.

The type IIB system with $N$ D9-branes and $N$ ghost
D9-branes also has a gauge anomaly which gets cancelled by
the Green-Schwarz mechanism.

\newsec{Heterotic Strings with Supergroup Gauge Symmetries}

\subsec{Heterotic World-Sheet from Type I/Heterotic Duality}

Since type I string theory with the $OSp(32+2n|2n)$ gauge group has
sixteen supersymmetries, it is well-motivated to consider its strong
coupling limit. We claim that it is given by the $OSp(32+2n|2n)$
heterotic string theory\foot{In this paper, we omit the difference
of supergroups analogous to the familiar one between $SO(32)$ and
$Spin(32)/{\IZ_2}$.} generalizing the well-known case of $n=0$,
i.e., the type I/heterotic duality \PW. The existence of this novel
heterotic string has already been mentioned in \TokunagaPJ\
recently.

We can derive its world-sheet theory from the type I side. A
D1-brane in the type I $OSp(32+2n|2n)$ string theory is dual to a
fundamental heterotic  string. We find  eight bosons $X^m\ \
(m=1,2,\ddd,8)$ and eight right-moving fermions $S^{\ti a}_{R}\ \
(\ti{a}=1,2,\ddd,8)$ as the transverse scalars and their
super-partners on the brane. In addition, we find $32+2n$
left-moving fermions $\lambda^i\ \ (i=1,2,\ddd,32+2n)$ and $2n$
left-moving bosons $\zeta^{\ti{\imath}}\ \ (\ti{\imath}=1,2,\ddd,
2n)$ from  1-9 strings.\foot{If we consider more than one D-strings,
there are gauge fields and their anomaly must be canceled \BSS. The
gauge anomaly is indeed canceled in our configuration of 9-branes.}
Here we used the fact that the open strings between the D1-brane and
the $n$ ghost D9-branes become ghost-like and have the wrong
spin-statistics relations. The world-sheet theory of the
$OSp(32+2n|2n)$ heterotic
 string in the light-cone Green-Schwarz formalism is thus
the $N=(0,1)$ conformal field theory with field content
\eqn\hetro{\eqalign{\hbox{left-moving}:&\ (X_L^m,\ \lambda^i_L, \
\zeta^{\ti{\imath}}_L),\ \ \ m=1,2,\ddd ,8;~ i=1,2,\ddd ,32+2n;~
\ti{\imath}=1,2,\ddd , 2n \cr \hbox{right-moving}:&\ (X_R^m,\
S^{\ti{a}}_{R}),\ \ \ \ \ \ \ \ \ m,\ti{a}=1,2,\ddd ,8.}}

The current algebra part of the heterotic world-sheet theory has
$32+2n$ spin-$1/2$ real fermions and $2n$ spin-$1/2$ real bosons,
which has the total central charge $c=16$.\foot{As we check in
appendix C, the central charge of the level-one current algebra for
$OSp(M|N)$ is $c=(N-M)/2$. Thus when $N-M=32$ the total central
charge vanishes in the left-moving sector: $c_L=10+16-26=0$.}
 Indeed, this is the free
field representation of the level-one $OSp(32+2n|2n)$
current algebra \GOW.

\subsec{Level-One $OSp(M|N)$ Current Algebra}

We now study the level-one $OSp(M|N)$ current ($N$ is even) algebra
since it is an essential building block of the heterotic
$OSp(32+2n|2n)$ string theory. For general aspects of current
algebras based on supergroups refer to, e.g., \GOW \BCMN.

We have the following OPEs for the $M$ free real fermions and $N$
real bosons (called symplectic bosons):
\eqn\opes{\lambda^i(z)\lambda^j(0)\sim \f{\delta^{ij}}{z}, \ \ \ \
\zeta^{\ti{\imath}}(z)\zeta^{\ti{\jmath}}(0)\sim
\f{-\eta^{\ti{\imath}\ti{\jmath}}} {z}.} We will raise the index of
symplectic bosons as $\zeta^{\ti{\imath}} =-\eta^{\ti{\imath}
\ti{\jmath}}\zeta_{\ti{\jmath}}$, where the anti-symmetric matrix
$\eta^{\ti{\imath}\ti{\jmath}} =iJ^{\ti{\imath}\ti{\jmath}}$ is
defined by (A.14) in appendix A.
 We define the bosonic currents
\eqn\boscur{J^a=\f{1}{2}t^a_{ij}\lambda^i\lambda^j, \ \ \
J^{\ti{a}}=\f{1}{2}t^{\ti{a}\ti{k}}_{\ti{\imath}}\zeta^{\ti{\imath}}\zeta_{\ti{k}},}
The bosonic generators $t^{a}$ and $t^{\ti{a}}$ belong to the
$SO(M)$ and $Sp(N)$ Lie algebras, respectively:
\eqn\constrle{t^a_{ij}=t^a_{ji},\ \ \
t^{\ti{a}i}_{j}=-\eta^{ik}t^{\ti{a}l}_k\eta_{lj}.} We also require
that $t^a$ and $t^{\ti a}$ are hermitian. In addition, there exist
fermionic currents\foot{Notice that the matrix $t^{\ap}$ is
bosonic.}
\eqn\fermioncur{J^{\ap}=t^{\ap}_{\ti{\imath}j}\zeta^{\ti{\imath}}\lambda^j.}

Then it is easy to show the currents \boscur\ and \fermioncur\ give
a representation of the $OSp(M|N)$ current algebra at level-one
\eqn\currentalsu{J^A(z)J^B(0) \sim \f{\f{1}{2}{\rm Str}[t^{A}t^{B}]}
{z^2}+\f{if^{AB}~_C~J^C(0)}{z},} where $A$ denotes all indices for
the adjoint representations of $OSp(32+2n|2n)$, i.e.
$A=\{a,\ti{a},\ap\}$. The structure constants $f^{AB}~_C$ are
defined by $[t^A,t^B]=if^{AB}~_C~ t^C$, where $[,]$ denotes an
anti-commutator if $A$ and $B$ are fermionic; otherwise it denotes a
commutator (see appendix A).

\subsec{Closed String Spectrum in the Heterotic $OSp(32+2n|2n)$
String}

In order to keep modular invariance, we impose the GSO projection
for the $(\lambda,\zeta)$ system in a way similar to the $SO(32)$
heterotic string \het \GreenSP. In the NS sector we have
anti-periodic boundary conditions
$\lambda^i(\tau,\sigma+2\pi)=-\lambda^i(\tau,\sigma)$ and
$\zeta^{\ti{\imath}}(\tau,\sigma+2\pi)=-\zeta^{\ti{\imath}}(\tau,\sigma)$.
The R sector is defined by  periodic boundary conditions. Let us
define $F_\lambda$ and $F_\zeta$ modulo 2 to be the operators that
count the numbers of $\lambda$ and $\zeta$ relative to the
appropriate ground state in each sector.\foot{ This ground state is
the $SL(2,\IC)$-invariant ground state in the NS sector, and
$|0\rangle_{\rm R}$ that will be defined later in the R sector.} The
GSO projection picks out states with $ (-1)^{F_\lambda+F_\zeta}=1 $
in both sectors. We define the torus partition function by \eqn\GSO{
Z={\rm Tr}{1+(-1)^{F_\lambda+F_\zeta}\over 2}(-1)^{F_\zeta}= {\rm
Tr}{(-1)^{F_\lambda}+(-1)^{F_\zeta}\over 2}, } where the trace is
over the left-moving NS and R sectors. The latter expression can be
written as a sum over the four spin structures of the torus, and
reduces to the torus partition function of the heterotic $SO(32)$
string due to cancellation between $2n$ fermions and $2n$ bosons.
Note that the cancellation requires the same periodicities of bosons
and fermions\foot{This also follows from the $\IZ_2$ residual gauge
symmetry on the D-string.}. These are the reasons for the
unconventional definition above and then the modular invariance is
obvious. The insertion of $(-1)^{F_\zeta}$ in the intermediate
expression implies that states with an odd number of $\zeta$
excitations have the opposite statistics to the usual heterotic states.
The total zero-point energies\foot{The physical state
condition is $(L_0+a)|{\rm phys}\lb=0$, where $a$ is the zero-point
energy.} of these sectors are given by $-1$ and $+1$ in the NS and R
sectors. Due to level matching with the right-moving sector, there
is no tachyon. The graviton multiplet arises in the standard way.
Massless gauge fields only come from the NS sector of the current
algebra. We find states corresponding to the $SO(32+2n)\times
Sp(2n)$ gauge bosons
\eqn\nsnststateso{\lambda^i_{-1/2}\lambda^j_{-1/2}|0\lb,\ \
\zeta^{\ti{\imath}}_{-1/2} \zeta^{\ti{\jmath}}_{-1/2}|0\lb,} as well
as their gauginos. Furthermore there exist gauge fields (and bosonic
gauginos)
\eqn\stateferso{\lambda^i_{-1/2}\zeta^{\ti{\jmath}}_{-1/2}|0\lb,}
which are fermionic because $(-1)^{F_\zeta}=-1$. Altogether, they
form the ten dimensional $OSp(32+2n|2n)$ super Yang-Mills multiplet.

Recently, it was found in \PolchinskiBG\
that there exists an open string in the
$SO(32)$ heterotic string theory.
One important consistency check for the existence of such an open string
was the cancellation of gauge non-invariant terms between the world-sheet
and spacetime.
Another was the conservation of degrees of freedom
flowing from the world-sheet to spacetime.
These checks go through for the $OSp(32+2n|2n)$ heterotic string
by replacing the traces by supertraces
as we did in subsection 3.2.

\vskip 0.5in

Another way to describe this heterotic string theory is to bosonize
the $(\lambda,\zeta)$ system. We can regard the $n$ copies of the
symplectic bosons $(\zeta^{2l-1},\zeta^{2l})\ \ \ (l=1,2,..,n)$ as
the $(\beta,\gamma)$ systems. Then we can apply the standard
bosonization procedure ($k=1,2,...,16+n)$
\eqn\bosoni{\eqalign{
&\lambda^{k}\equiv\f{1}{\s{2}}(\lambda^{2k-1}
+i\lambda^{2k})=e^{i\vp^{k}},\ \
\bar{\lambda}^{k}\equiv\f{1}{\s{2}}(\lambda^{2k-1}
-i\lambda^{2k})=e^{-i\vp^{k}},\cr
&\zeta^l\equiv{1\over\sqrt{2}}(\zeta^{2l-1}+i\zeta^{2l})= e^{i\tvp^{l}}\de\xi^l,\ \
\bar{\zeta}^l\equiv{1\over\sqrt{2}}(\zeta^{2l-1}-i\zeta^{2l})
=e^{-i\tvp^{l}}\eta^l,}} where the OPEs are \eqn\opebpos{
\vp^{k}(z)\vp^{k'}(0)\sim -\delta^{kk'}\log z, \ \ \
\tvp^{l}(z)\tvp^{l'}(0)\sim \delta^{ll'}\log z,\ \ \
\eta^l(z)\xi^{l'}(0)\sim \f{\delta^{ll'}}{z}.}

We can represent the Cartan subalgebra generators $H_{m}\ \ (m=1,2,\ddd,16+2n)$
of $OSp(32+2n|2n)$ as the following $16+2n$ currents
\eqn\cartan{\eqalign{& H_{l}=\zeta^l\bar{\zeta}^l=-i\de\tvp^l,
\cr &H_{n+k}=\lambda^k\bar{\lambda^k}=i\de\vp^k.}} The vertex
operator of the form ($V_{\rm oscillator}$ denotes the part made of
oscillators of $\vp,\tvp,\eta$ and $\xi$)
\eqn\primary{V_{\vec{\ap}}=\exp\Big[i\sum_{k=1}^{16+n}\ap_{n+k}\vp^k
+i\sum_{l=1}^{n}\ap_{l}\tvp^l\Big]\cdot V_{\rm oscillator},} possesses
the weight eigenvalues $H_{m}=\ap_m$ and its conformal dimension is
\eqn\confdim{\Delta=-\f{1}{2}\sum_{m=1}^{n}(\ap_m)^2+\f{1}{2}
\sum_{m=n+1}^{16+2n}(\ap_{m})^2+\Delta_{\rm oscillator}\equiv
\f{1}{2}({\ap},{\ap})+\Delta_{\rm oscillator}.} Notice that the
conformal dimension $\Delta_{\rm oscillator}$ for the oscillator part is
always a non-negative integer.

The $16+2n$ simple roots of $OSp(32+2n|2n)$ (see the Dynkin diagram
in Fig.6) are given by the following operators (we omit cocycle
factors) \eqn\simpleroots{\eqalign{&
J^{\ap_l}=e^{i\tvp^l-i\tvp^{l+1}}\de\xi^{l}\eta^{l+1}\ \ \ \
(l=1,2,\ddd,n-1), \cr &J^{\ap_n}=e^{i\tvp^n-i\vp^1}\de\xi^n, \cr &
J^{\ap_{n+k}}=e^{i\vp^k-i\vp^{k+1}}\ \ \ \ (k=1,2,\ddd,15+n),\cr
&J^{\ap_{16+2n}}=e^{i\vp^{15+n}+i\vp^{16+n}}.}} Notice that $\ap_n$
is the only fermionic simple root and the others are all bosonic. We
find the inner products (called the symmetric Cartan matrix\foot{
Distinguish this from the ordinary Cartan matrix $A_{mm'}$ whose
diagonal part is always $2$ or $0$.} $A'_{mm'}$)
\eqn\catranmat{\eqalign{({\ap}_m, {\ap}_{m'})&=-2\ \ \ {\rm if}\
1\leq m=m'\leq n-1,\cr &=2 \ \ \ {\rm if}\ n+1\leq m=m'\leq
16+2n,\cr &=1\ \ \ {\rm if}\ \ap_m\ {\rm and}\ \ap_{m'}~ {\rm are~
adjacent~ in~ the~ Dynkin~ diagram~ and}\ 1\leq m,m'\leq n ,\cr
&=-1\ \ \ {\rm if}\ \ap_m\ {\rm and}\ \ap_{m'}~ {\rm are~ adjacent~
in~ the~ Dynkin~ diagram~ and}\ n\leq m,m'\leq 16+2n,\cr &=0 \ \ \
{\rm in~ other~ cases}. }} From \simpleroots, we see that the root
lattice of $OSp(32+2n|2n)$ is given by \eqn\rootlattice{\eqalign{
\Gamma_{\rm root}=\{ (n_1,...,n_{16+2n})\in \IZ^{16+2n}|\sum_{l=1}^n
n_l+\sum_{k=1}^{16+n} n_{n+k}\in 2\IZ\} }}

The weights of the `spinor' representation can be found as follows.
In terms of the $\lambda\zeta$ system, the fields corresponding to
simple roots are \eqn\simplerootsfermi{\eqalign{ & J^{\ap_l}\propto
\bar{\zeta}^{l}\zeta^{l}  \ \ \ \ (l=1,2,\ddd,n-1), \cr
&J^{\ap_n}\propto\bar{\lambda}^1\zeta^{2n-1}, \cr
&J^{\ap_{n+k}}\propto\bar{\lambda}^{k+1}\lambda^k\ \ \ \
(k=1,2,\ddd,15+n),\cr
&J^{\ap_{16+2n}}\propto\lambda^{15+n}\lambda^{16+n}. }} As in the
$SO(32)$ case, the GSO projected R-ground states furnish a `spinor'
representation. Let us define $|0\rangle_R$ to be the ground state
annihilated by the zero-modes of $\lambda^{k}$ and $\zeta^l$. Since
the simple root operators contain at least one of these,
$|0\rangle_R$ is the highest weight state in one of the irreducible
spinor representations. This state has eigenvalues (highest weight)
\eqn\spinorhighestweight{ ((-1/2)^n,(1/2)^{16+n}) } of the Cartan
generators. Thus an element of the weight lattice $\Gamma_{\rm
spinor}$
 for this spinor representation is the sum of \spinorhighestweight\ and a vector
in $\Gamma_{\rm root}$.
The Narain lattice for the bosonized description of the
 $OSp(32+2n|2n)$ heterotic
string is the sum of the two lattices:
\eqn\osplattice{
\Gamma_{n,16+n}=\Gamma_{\rm root}\cup \Gamma_{\rm spinor}.
}
The inner product of $q,q'\in \Gamma_{n,16+n}$  is
\eqn\innerprod{
(q,q')= -\sum_{l=1}^n q_l q'_l +\sum_{k=1}^{16+n} q_{n+k}q'_{n+k}.
}
as we defined in \confdim.
It is easy to check that the lattice
$\Gamma_{n,16+n}$ is even with respect to this inner product
as required by the level matching condition and locality of vertex operators.
Although the torus partition function has contributions from fermions $\eta,\xi$,
we expect that modular invariance requires self-duality of the lattice.
We have checked that the lattice $\Gamma_{n,16+n}$ is indeed self-dual.

\subsec{Heterotic String Based on the $E_8\times E_8$-like Supergroup}

We can define another heterotic string by imposing double GSO
projections as we do to define the $E_8\times E_8$ heterotic string
\het\ (see also \TokunagaPJ\ for an earlier discussion). We expect that
this leads to another supergroup extension of heterotic string
theory in ten dimensions. Let $n$ be even.
 We divide the free fields $\lambda^i$ and
$\zeta^{\ti{\imath}}$ in the previous subsection into two groups:
\eqn\groupss{\eqalign{ (\lambda^i,\zeta^{\ti{\imath}}) :& \ \ \
i=1,2,\ddd,16+n;\ \ \ti{\imath}=1,2,\ddd,n,\cr
(\lambda^{i'},\zeta^{\ti{\imath}'}) :& \ \ \
i'=17+n,18+n,\ddd,32+2n; \ \ \ti{\imath}'=n+1,n+2,\ddd,2n.}} After
taking GSO projections \GSO\ separately on these theories, we have
the four left-moving sectors $({\rm NS},{\rm NS})$, $({\rm NS},{\rm
R})$, $({\rm R},{\rm NS})$ and $({\rm R},{\rm R})$. As in the $OSp$
case, the $({\rm NS},{\rm NS})$ sector has the total zero-point
energy $-1$ and thus the massless gauge bosons for $SO(16+n)\times
SO(16+n)\times Sp(n)\times Sp(n)$ are
\eqn\nsnststate{\lambda^i_{-1/2}\lambda^j_{-1/2}|0\lb,\ \
\lambda^{i'}_{-1/2}\lambda^{j'}_{-1/2}|0\lb,\ \
\zeta^{\ti{\imath}}_{-1/2} \zeta^{\ti{\jmath}}_{-1/2}|0\lb,\ \
\zeta^{\ti{\imath'}}_{-1/2} \zeta^{\ti{\jmath'}}_{-1/2}|0\lb.} There
also exist fermionic gauge fields
\eqn\statefer{\lambda^{i}_{-1/2}\zeta^{\ti{\jmath}}_{-1/2}|0\lb,\ \
\lambda^{i'}_{-1/2}\zeta^{\ti \jmath'}_{-1/2}|0\lb.}

Furthermore, we have other massless states from the $({\rm NS},{\rm
R})$ and $({\rm R},{\rm NS})$ sectors. Since the zero-point energy
vanishes, the ground states give rise to massless fields. The
degeneracy of ground states comes from the fermionic zero-modes
$(\lambda^i_0,\lambda^{i'}_0)$ and bosonic zero-modes
$(\zeta^{\ti{\imath}}_{0},\zeta^{\ti{\imath'}}_{0})$. The former, as
is familiar in the ordinary heterotic string theory, leads to the
spinor representations with dimension $2^{7+\f{n}{2}}$ (assuming $n$
is even) of either of the two $SO(16+n)$s. The latter is a novel
ingredient in this kind of heterotic string. Indeed it generates
infinitely many massless modes because the bosonic zero-modes
constitute the Heisenberg algebra (or equivalently the spinor
representation of the metaplectic group).

In the ordinary $E_8\times E_8$ heterotic  string,  the
248 dimensional adjoint representation of one
$E_8$ is obtained by combining the 120
dimensional adjoint representation of one $SO(16)$ in the
$({\rm NS},{\rm NS})$ sector and the $2^7=$128 dimensional
spinor representation of the same $SO(16)$. In our heterotic
 string we can see that the
massless gauge bosons and fermions belong to  two copies of an
infinite dimensional Lie superalgebra. We call this superalgebra
$E(8+{n\over 2},{n\over 2})$ since it includes the $E_8$ algebra.
What we have discussed is the heterotic $E(8+{n\over 2},{n\over 2})
\times E(8+{n\over 2},{n\over 2})$
 string.

We found that
$E(8+{n\over 2},{n\over 2})$ is infinite dimensional.
This fact seems to be consistent with the known mathematical
fact:
there is no finite dimensional Lie superalgebra which is a
counterpart for the $E_n$ Lie algebra. The only examples of
exceptional Lie superalgebras are called $G(3)$ and $F(4)$,
whose
bosonic parts are $G_2\times SU(2)$ and $SO(7)\times SU(2)$,
respectively. Indeed, if we try to extend the $E_8$ algebra by adding
fermionic roots, we find that the Cartan matrix
ceases to be positive definite. Thus the superalgebra becomes
infinite dimensional (this is called an indefinite superalgebra).

The Narain lattice for the $E(8+{n\over 2},{n\over 2})
\times E(8+{n\over 2},{n\over 2})$
 heterotic string is $\Gamma_{{n\over 2},8+{n\over 2}}
 \times \Gamma_{{n\over 2},8+{n\over 2}}$,
 where $\Gamma_{{n\over 2},8+{n\over 2}}$ is the root lattice of
$E(8+{n\over 2},{n\over 2})$ and is defined in the same
 way as $\Gamma_{n,16+n}$:
An element of $\Gamma_{{n\over 2},8+{n\over 2}}$ is an integer
vector $(n_1,...,n_{8+n})$ such that $\sum_{a} n_a$ is even, or the
sum of such a vector and $((-1/2)^{n/2},(1/2)^{8+n/2})$. It is
equipped with the $(-^{n/2},+^{8+n/2})$-signature metric.

We can discuss the strong coupling limit of this heterotic string
theory. The product form of the gauge supergroup suggests the
Horava-Witten type duality \HoWi. In other words, we expect that
this ten dimensional string theory is dual to the M-theory on
$S^1/\IZ_2$. This $\IZ_2$ projection preserves sixteen
supersymmetries. Each of two fixed planes will provide the
$E(8+{n\over 2},{n\over 2})$ gauge theory. It would be interesting
to see this from  the anomaly cancellation argument in eleven
dimensional supergravity.

Finally we would like to mention a subtlety that appears when we
consider the spinor representation for the $OSp$ supergroups. Such a
state appears in the massless states of the heterotic $E(8+{n\over
2},{n\over 2}) \times E(8+{n\over 2},{n\over 2})$ string as we have
seen, and also in the massive states of the heterotic
$OSp(32+2n|2n)$ string. The corresponding vetex operators can be
constructed via the bosonization \bosoni\ of the $(\beta,\gamma)$
system. To maintain the modular invariance, we need to pick up a
state with a definite picture as in the superconformal ghosts sector
of the ordinary superstrings. When we consider an OPE between two
different R-sector operators we encounter an vertex operator with a
different picture. Then we need to identify states with different
pictures as we do by the picture changing operation in ordinary
superstrings. We leave the details of this operation in our case for
a future problem.

\subsec{Toroidal Compactification}

One can consider the heterotic $OSp(32+2n|2n)$ or $E(8+{n\over
2},{n\over 2})\times E(8+{n\over 2},{n\over 2})$ string theory
compactified on $T^d$. As in the usual heterotic string theory, it
is convenient to use the bosonized description. We use coordinates
such that the radius of each circle is $R$.

Turn on constant metric $g_{\mu\nu}~(\mu,\nu=1,...,d)$, $B$-field
$B_{\mu\nu}$, and Wilson lines $A^l_\mu $ ($l=1,...,n$) and
$A^k_\mu$ ($k=n+1,2,...,16+2n$). As is shown in appendix D, the
momenta for this system are given by \eqn\momlattice{\eqalign{
k_{L\mu}&={n_\mu \over R} + {w^\nu  R\over \alpha'}(g_{\mu\nu}+
B_{\mu\nu})-q_l A^l_\mu-q_k A^k_\mu - {w^\nu R\over 2} ( -A^l_\nu
A^l_\mu +A^k_\nu  A^k_\mu),\cr k_{l}&=(q_l-w^\mu  R A^l_\mu
)\sqrt{2\over \alpha'},\cr k_{k}&=(q_k+w^\mu  R A^k_\mu
)\sqrt{2\over \alpha'},\cr k_{R\mu}&= {n_\mu \over R}+ {w^\nu  R\over
\alpha'} (-g_{\mu\nu}+B_{\mu\nu})-q_lA^l_\mu-q_k A^k_\mu - {w^\nu
R\over 2}(-A^l_\nu  A^l_\mu+A^k_\nu A^k_\mu  ), }} generalizing the
results for ordinary heterotic strings \refs{\PolII}.
Here $(q_l, q_k)$ is a point
in the lattice $\Gamma_{n,16+n}$ or $\Gamma_{{n\over 2},8+{n\over
2}}\times \Gamma_{{n\over 2},8+{n\over 2}}$
 that defines the heterotic string theory.
$n_\mu$ and $w^\mu$ are arbitrary integers
representing the momentum and the winding number along the $x^\mu$ direction.

Let us define
$$
l:=(\sqrt{\alpha'\over 2}k_{L\mu}, \sqrt{\alpha'\over 2}
k_l, \sqrt{\alpha'\over 2}k_k, \sqrt{\alpha'\over 2}k_{R\mu}).
$$
Then the level matching condition is
$$
0=L_0-\bar{L}_0={1\over 2}l\circ l+N-\bar{N}-1.
$$
Here $N$ and $\bar{N}$ arise from oscillator excitations and take integer values.
We have defined the metric on the momentum lattice by
$$\eqalign{
l\circ l'&={\alpha'\over 2}(g^{\mu\nu} k_{L\mu}
k'_{L\nu}-k_l{k'}_l+k_k {k'}_k -g^{\mu\nu}k_{R\mu} k'_{R\nu})\cr
&=n_\mu w'^\mu+w^\mu n'_\mu -q_l q'_l+q_k q'_k. }$$ We see that the
level matching condition is satisfied because the  lattice
$\Gamma_{n,16+n}$ or $\Gamma_{{n\over 2},8+{n\over 2}}$ of
$(q_l,q_k)$ is even. It is clear that the moduli space of the
lattices for this toroidal compactification is given by
\eqn\narainl{ SO(16+d+n,d+n;\IZ)\Bigl\backslash SO(16+d+n,d+n;{\bf
R})\Bigl/ SO(16+d+n,n;{\bf R})\times SO(d,{\bf R}).} The moduli are
the metric, $B$-field, and Wilson lines.

As is well-known the $SO(32)$ and $E_8\times E_8$ heterotic strings
become equivalent upon $S^1$ compactification by choosing
appropriate Wilson lines and radii.
This equivalence extends to the $OSp(32+2n|2n)$ and
$E(8+{n\over 2},{n\over 2})\times E(8+{n\over 2},{n\over 2})$ strings.
Let us turn on Wilson lines
$$
(A^l, A^k)=((1/2R)^{n/2},0^{n/2},(1/2R)^{8+n/2},0^{8+n/2})
$$
on the $OSp$ side and
$$
(A^l, A^k)=(0^n,1/R,0^{7+n/2},1/R,0^{7+n/2})
$$
on the $E\times E$ side.
It is cumbersome but straightforward to show that
the spectrum \momlattice\ on the $OSp$ side is exchanged with
that on the $E\times E$ side via $R\rightarrow \alpha'/2R$,
$(k_L,k_R)\rightarrow (k_L,-k_R)$.

\newsec{Type II-like Closed Superstrings with $U(n|n)$ Supergroup Gauge Symmetries}

In the previous section, we constructed the world-sheet of the
heterotic $OSp(32+2n|2n)$ string from the strong coupling limit of a
D1-brane in the type I $OSp(32+2n|2n)$ string, generalizing the
typeI/heterotic duality. In this section, we consider the type IIB
S-duality in the same spirit and ask what the world-sheet
description is for the S-dual of the system involving $n$ D9-branes
and $n$ ghost D9-branes. In other words we study type IIB string
with $n$ NS9-brane and $n$ ghost NS9-branes. We will be able to
construct a IIB-like superstring world-sheet which leads to the
$U(n|n)$ gauge symmetry.

\subsec{Superstring World-Sheet from S-Duality}

Consider the world-sheet of a D-string in the background of $n$
D9-branes and $n$ ghost D9-branes. We find eight transverse scalars
$X^m$ ($m=1,2,\cdots,8$), a non-dynamical gauge field $A_\mu$, eight
left-moving fermions $S^a_L$ ($a=1,2,\cdots, 8$), and eight
right-moving fermions $S^{\tilde a}_R$ ($\tilde{a}=1,2,\cdots, 8$)
from the 1-1 string. $S^a_L$ and $S^{\tilde{a}}_R$ transform in the
spinor representations of opposite chiralities\foot{Here we imposed
the Dirac equation.}. These fields from the 1-1 string match the
massless excitations of the type IIB fundamental string \WittenIM.
We have new ingredients due to the 9-branes. From the 1-9 strings,
we find left-moving fermions $\lambda^i$ ($i=1,...,n$) and
left-moving (ghost) bosons $\zeta^{\tilde{\imath}}$
($\tilde{\imath}=1,...,n$). The 9-1 strings give the conjugate
fields $\bar{\lambda}_i$ and $\bar{\zeta}_{\tilde{\imath}}$.
These spin-${1\over 2}$ fermions and bosons behave just like those
in \bosoni.
They furnish a representation of the level-one
current algebra $OSp(2n|2n)_{k=1}$, which
will reduce to $U(n|n)_{k=1}$ as we will see below.

We also need to take into account the effects of the gauge field
$A_\mu$ on the D1-brane. Let us consider a compactification on a
circle of radius 1 in appropriate coordinates and assume that the
D-string is wrapping the circle. It is well-known that the flux
$F_{\tau\sigma}$ measures the fundamental string charge on the
D-string. Since we are interested in the pure D-string (e.g. the
perturbative F-string in the dual side), it is natural to consider
the sector with $F_{\tau\sigma}=0$. Then the path-integral for
$A_\mu$ after gauge fixing is over the constant Wilson line
$A_\sigma$. Since $A_\mu$ couples with the $U(1)$ current
\eqn\uonecr{J=-\bar{\lambda}_i \lambda^i-\bar{\zeta}_{\ti\imath}
\zeta^{\ti\imath},} the integration over constant $A_\sigma$ forces
the $U(1)$ charge $J_0$ to vanish: \eqn\uonec{J_0\equiv\int d\sigma
J(\sigma)=0.} The restriction to the zero-charge sector reduces the
current algebra\foot{If we fully gauge the $U(1)$ without imposing
$F_{\tau\sigma}=0$, we get the current algebra $PSU(n|n)_{k=1}$.} to
$U(n|n)_{k=1}$ (see \BCMN\ for the properties of this current
algebra).

Let us combine the fields as
$(\Lambda^I)=(\lambda^i,\zeta^{\tilde{\imath}}),
(\bar{\Lambda}_I)=(\bar{\lambda}_i,\bar{\zeta}_{\tilde{\imath}})$.
Turning on a Wilson line along the $S^1$ changes the periodicities
to $\Lambda^I(\sigma+2\pi)=e^{2\pi i\nu}\Lambda^I(\sigma),~
\bar{\Lambda}_I(\sigma+2\pi)=e^{-2\pi i\nu}\bar{\Lambda}_I(\sigma)$
for some real $\nu$. Just as the $\IZ_2$ gauge symmetry produces the
R- and NS-sectors of the heterotic string in the type I/heterotic
duality, the $U(1)$ gauge symmetry instructs us to integrate over
$\nu$ from 0 to 1. This integration is however trivial because as
long as we look at sectors where the $U(1)$ charge $J_0$
vanishes,\foot{ The requirement $J_0|{\rm state}\rangle=0$ can also
be regarded as the analog of the GSO projection in the heterotic
string case.} the Hilbert space is independent of $\nu$\foot{For
example, the $J_0$ condition excludes
$\bar{\lambda}_{i0}\bar{\lambda}_{j0}|0\rangle_{\rm R}$ that has no
counterpart in the NS sector.}. We choose to work in the NS-sector
($\nu=1/2$) in what follows.

\subsec{Closed Superstrings with $U(n|n)$ Gauge Symmetries}

We are led to consider a superstring whose world-sheet theory is
described in the NSR-formulation by \eqn\IIcontent{\eqalign{
\hbox{left-moving}&:(X^\mu_L,\psi^\mu_L) \times U(n|n)_{k=1},\cr
\hbox{right-moving}&:(X^\mu_R,\psi^\mu_R), }} with the $U(1)$
projection \uonec. As is usual in the type II string, we also have
the $(b,c)$ and $(\beta,\gamma)$ ghosts.

Now we briefly study the new superstring we have just discovered.
In the $(0,0)$-picture, the gauge fields are represented
by the heterotic-like vertex operators \eqn\vertex{
J^\alpha(z)\bar{\partial}X^\mu(\bar{z}) e^{ik\cdot X}+\cdots, }
where $J^\alpha=\bar{\Lambda}_I(t^\alpha)^I_{~J}\Lambda^J$ are
the $U(n|n)_{k=1}$ currents.
The modular invariance again follows from the modular
invariance of the usual type IIB superstring
due to the cancellation between $\lambda$s and $\zeta$s.
Note that the partition function involves integration over
the periodicity $\nu$ in the $\sigma$ direction
as well as the periodicity $\nu'$ in the $\tau$-direction.
The $\nu'$ integral is equivalent to the $J_0$ condition.
Both integrals are trivial because the integrand is one.

Clearly, this superstring with $U(n|n)$ gauge symmetry is equivalent
to the ordinary type IIB superstring as long as the amplitude
involves only the external particles that are present in the usual
theory. This is because the system with $n$ D-branes and $n$ ghost
D-branes is equivalent to the one without branes, and our theory is
S-dual to such a system. This equivalence holds perturbatively
because of the fact that a $U(n|n)$ gauge theory is trivial as we
saw in subsection 3.1. This equivalence is no longer true when we
compactify the string theory on a circle and turn on generic Wilson
lines. For example, it is possible to turn on the Wilson lines so
that the interactions between the  NS9-branes and the ghost NS9-branes,
which carry $U(n)$ gauge groups, become weak.
This may be a useful model to investigate NS9-branes.

The existence of a gauge multiplet implies that the supersymmetry is
superficially broken from 32 supercharges to 16 supercharges, though
the system sitting at the vacuum is actually equivalent to the type
IIB superstring. We expect that the construction here extends to
IIA-like superstrings via T-duality. We would like to come back to
more details of these new string theories in another publication.

\newsec{Lie Superalgebras from 7-Brane Configurations}

Type I $SO(32)$ string theory on $T^2$ is equivalent to the type IIB
string theory on $T^2/{\IZ}_2$ with four orientifold 7-branes
(O7-branes) located at each fixed point via the T-duality. There are
sixteen D7-branes allowed so that the tadpoles are canceled. This
string theory is known to be non-perturbatively described by
F-theory compactified on a specific elliptically fibered K3 surface
\VF \Sen. In the latter description the presence of D7-branes and
O7-branes is equivalent to the existence of singular fibers in the
K3 surface.

If we consider a probe D3-brane near an O7-brane, its low energy
theory is given by the 4D $N=2$ $SU(2)$ super Yang-Mills theory with
various flavor symmetries depending on the configurations of D7-branes. For
example, if we have $N_f$ D7-branes located at the orientifold, then
the flavor symmetry is $D_{N_f}=SO(2N_f)$. The quarks are realized
as the $(p,q)$ strings or their string junctions between the
D3-brane and various 7-branes. It is even possible to realize the
$E_{6,7,8}$ flavor symmetries. In summary, we can realize $A_n,D_n$
and $E_{6,7,8}$ (i.e. all simply laced Lie algebras) symmetries by
this method (refer to \DZ\ and references therein). These models of
7-branes and string junctions provide us with a visual way of understanding
various enhanced symmetries in string theory.

\fig{The $SU(M|N)$ Dynkin diagram from a 7-brane configuration in type
IIB string theory theory. A white box denotes a D7-brane and a black one a
ghost D7-brane. A line with an arrow between 7-branes represents an
F-string, which corresponds to one of the simple roots
$\ap_1,\ap_2,\ddd,\ap_{N+M-1}$. The $N$-th F-string becomes
fermionic because it stretches between a 7-brane and a ghost 7-brane
agreeing with the fact that $\ap_N$ is a fermionic simple
root.}{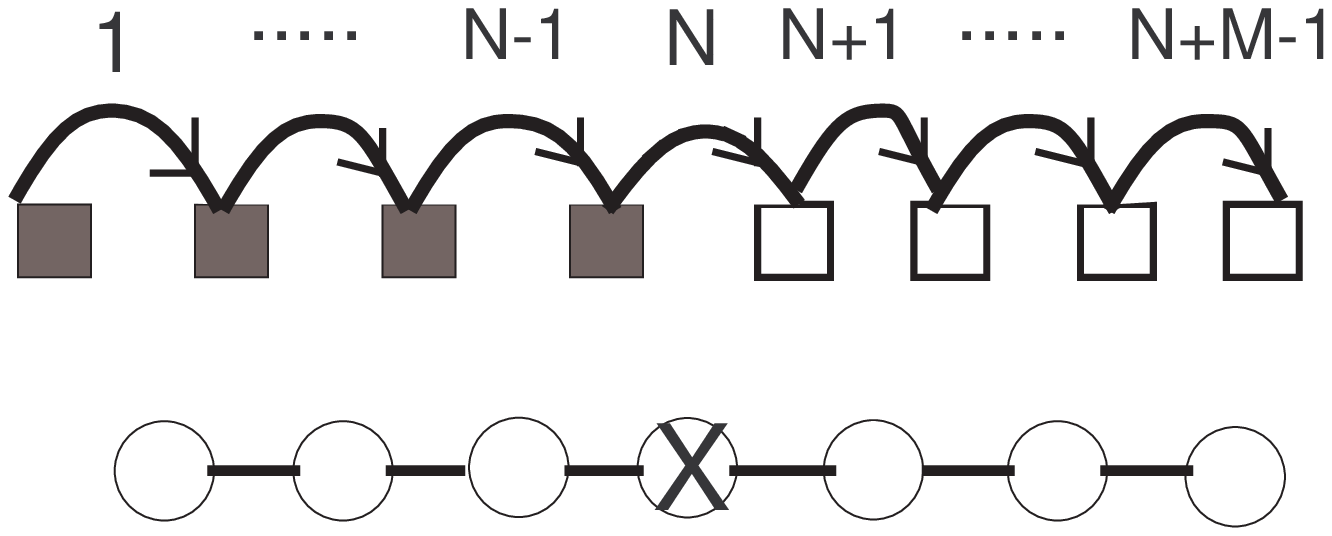}{2.2truein}

In this setup of 7-branes in type IIB string, we can again introduce
ghost D7-branes without breaking the sixteen supersymmetries. Notice
that a ghost 7-brane possesses the opposite monodromy $\tau(z)\sim
-{1\over 2\pi i}\log z$ of the dilaton-axion relative to the
ordinary 7-brane. Let us start with  $M$ D7-branes and $N$ ghost
D7-branes. Then we can see that the open fundamental strings between
them lead to the adjoint representations of the Lie superalgebra
$SU(M|N)$. Indeed we can pick up simple roots naturally from the
brane configuration and find the structure of the Dynkin diagram as
in Fig.2. One important new ingredient is that in the superalgebra
one of the simple roots is fermionic and indeed it is realized as an
open string between a D7-brane and a ghost D7-brane.

\fig{The $OSp(2M|2N)$ Dynkin diagram from a 7-brane configuration in
type IIB string. The white box denotes a D7-brane (or $(1,0)$
7-brane) and the black one a ghost D7-brane. The circles B and C
are $(1,1)$ and $(1,-1)$ 7-branes, respectively. The $N$-th F-string
becomes fermionic because it stretches between a 7-brane and a ghost
7-brane, corresponding to the fermionic simple root $\ap_{N}$. The
final root $\ap_{N+M}$ is the string junction made of $(2,0)$,
$(1,1)$ and $(1,-1)$ strings.}{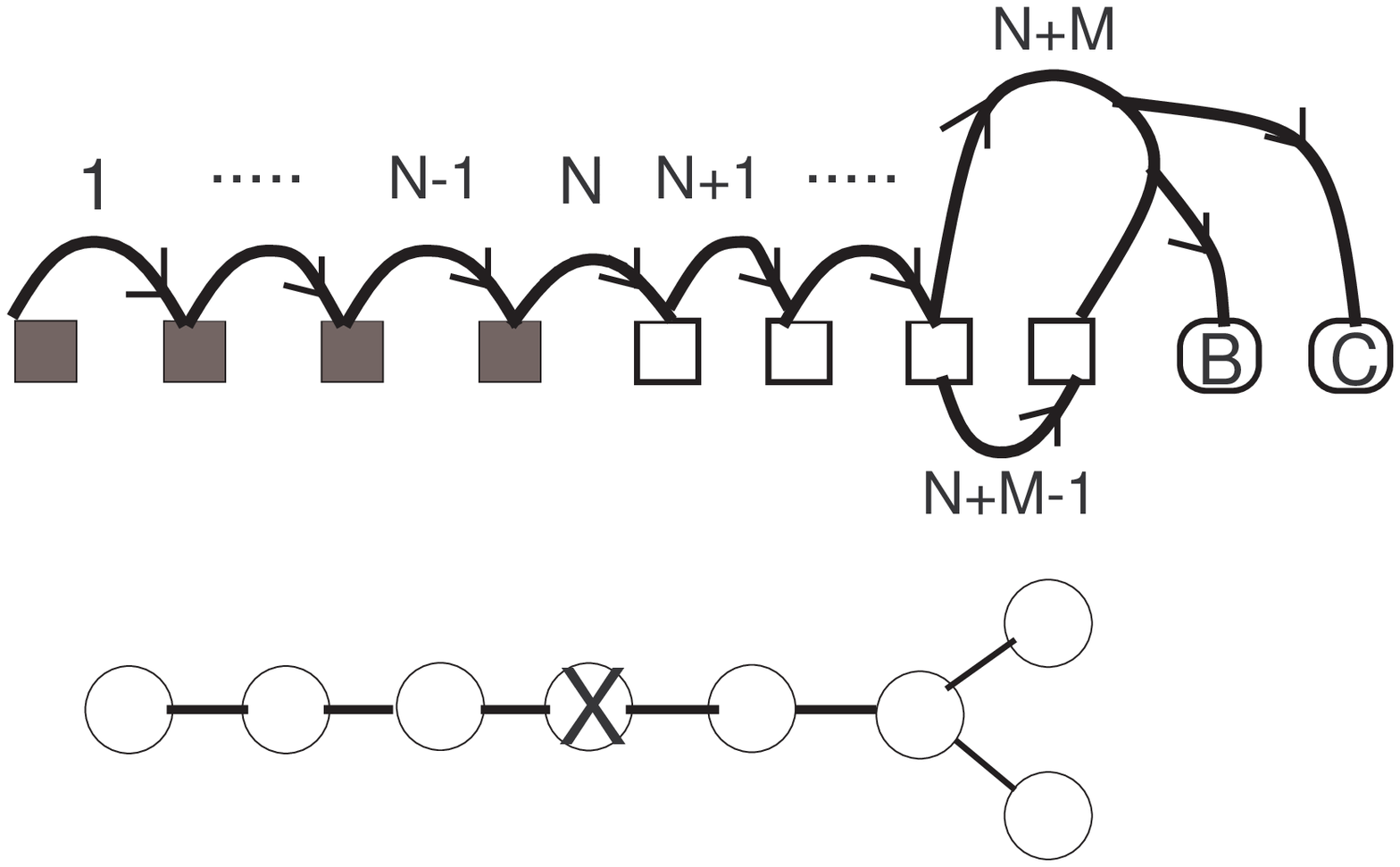}{2.5truein}

Furthermore we can add an O7-brane in this 7-brane configuration.
Then the gauge group enhances into $OSp(2M|2N)$.
 The O7-brane can be regarded as
a bound state of one $(1,-1)$ and one $(1,1)$ 7-brane \DZ. Then we
can express the simple roots and the Dynkin diagram of $OSp(2M|2N)$
in terms of string junctions as in Fig.3. Even though the $SO(2M)$
part of the bosonic subgroup is manifest in Fig.3, the $Sp(2N)$
symmetry is not obvious. In fact, we can find the string junction
corresponding to the long root with length-squared four which is
typical in $Sp(2N)$ as in Fig.4. In terms of simple roots of the
$OSp$ algebra the long root $\ap_{\rm long}$ is given by
\eqn\longroot{\ap_{\rm long}=2(\ap_N+\ap_{N+1}+\ddd+\ap_{N+M-2})
+\ap_{N+M-1}+\ap_{N+M}.} Then one may worry that the long root may
contradict with the standard BPS condition $Q^2=-(\ap,\ap)\geq -2$,
where $Q^2$ is the intersection number of the string junction. In
fact, this was the reason why we can not realize the non-simply
laced symmetry\foot{However, it is known that the $Sp(N)$ gauge
group can be realized in the presence of discrete fluxes on K3
surfaces \SPJ, which is dual to the CHL heterotic string theory
\CHL.} in the F-theory on K3.

However, in our model with ghost D7-branes, the long root is
actually BPS. First of all, the intersection number should be
defined with a minus sign for the open string which starts or ends
at ghost branes. Thus we should have $(\ap_{m},\ap_{m})=-2$ for
$m=1,2,\ddd,N-1$, $(\ap_{N},\ap_{N})=0$, and $(\ap_{m},\ap_{m})=2$
for $m=N+1,2,\ddd,N+M$ (see also \catranmat). In these rules, the
length-squared of the long root should be regarded as $(\ap_{\rm
long},\ap_{\rm long})=-4$. Moreover, the BPS condition also becomes
more complicated than the one in the ordinary setup. This can be
easily understood from the heterotic dual viewpoint. Consider the
heterotic $OSp(32+2n|2n)$ string on $T^2$. The BPS condition
requires that the right-moving sector is in the ground state. Thus
from \confdim\ we immediately find \eqn\BPS{(\ap_{\rm long},\ap_{\rm
long})\leq 2-2\Delta_{\rm oscillator},} where we assumed that there
are no momenta in the compactified directions. When the equality is
saturated, the mode becomes massless. For the long root $(\ap_{\rm
long},\ap_{\rm long})=-4$, the left-moving part of the vertex
operator looks like \eqn\vertex{J^{\ap_{\rm
long}}=e^{2i\tvp_N}\de\xi\de^2\xi.} Thus this includes the
oscillator excitation of $\Delta_{\rm oscillator}=3$ and it indeed
becomes massless.

\fig{The description of the long root $\ap_{\rm long}^2=4$ in
$Sp(2N)$ in terms of a string junction. Even though a long root is
not a BPS junction in ordinary string theory, it becomes BPS in our
setup which includes ghost 7-branes. This is not a simple root in
the superalgebra $OSp(2M|2N)$ and thus did not show up in
Fig.3.}{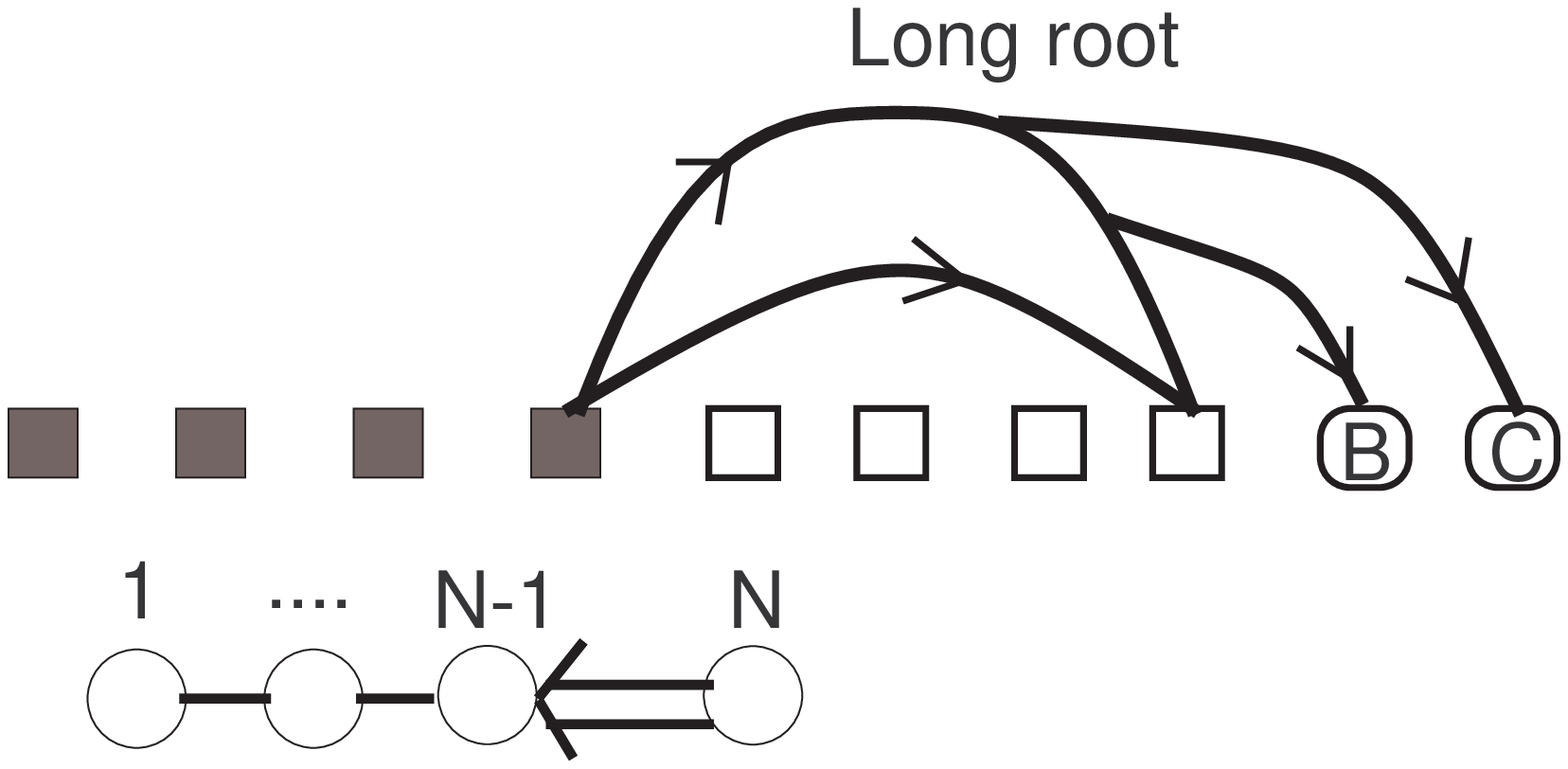}{2.2truein}

It is possible to obtain a superalgebra counterpart of $E_n$ by
adding ghost D7-branes in the $E_n$ 7-brane configuration. Such
superalgebras are, as we have seen from the heterotic string
viewpoint, infinite dimensional because the Cartan matrix $A_{ij}$
is no longer positive definite. Such algebras are called indefinite
superalgebras. One simple example is obtained from the $D_5$ Dynkin
diagram (=7-brane configuration) by adding a fermionic node (=a
ghost D7-brane) so that the Dynkin diagram becomes similar to $E_6$
(see e.g. \FS). It will also be interesting to see if we can obtain
the affine Lie superalgebras from 7-brane configurations.

Finally, we discuss the F-theory interpretation. When we separate
ghost 7-branes from D7-branes, the value of Im$\tau$ ($\tau\equiv
ie^{-\phi}+\chi$) is negative near a ghost brane. This is not
possible if we identify $\tau$ with the period of the torus fiber in
the elliptically fibered K3 surface. This suggests that we need to
consider F-theory on a sort of generalized K3 surface. We expect
that such a manifold would be a supermanifold with complex
superdimension two\foot{Some of earlier works on supermanifolds are,
e.g.,  \Sethi \ASchwarz \TokunagaPJ. More recently supermanifolds
have been discussed in the topological string theory on twistor
spaces in \Witten \Aganagic \Seki.}. We encounter a similar
situation when we examine the type IIA/heterotic duality as we
discuss in the final section.

\newsec{Discussions}

\subsec{Isolated Ghost D-branes}

When $M$ ghost D-branes are on top of $N\geq M$ ordinary D-branes
with trivial gauge backgrounds, the effects of ghost D-branes are
completely canceled. It is clear that there is no physical pathology
in this system. The physical interpretation is subtler when some
background fields are turned on and ghost D-branes are not canceled
by ordinary branes. For example, when ghost D-branes are separated
from ordinary D-branes, the system has negative tension objects on
which ghost fields appear. The Born-Infeld analysis seems to
indicate that the minimum energy configuration is such that the
ghost D-branes are moving at the speed of light and the D-branes are
static. This configuration is non-supersymmetric and may decay into
some other background.

On the other hand, we have seen several interesting properties that
may turn ghost D-branes into a useful notion in string theory. First
of all, they preserve the same supercharges as ordinary D-branes, so
the combined system is BPS. Second, in the compactified heterotic
string dual discussed in section 4, we found the degrees of freedom
corresponding to turning on the Wilson line. They are T-dual to
moving ghost D8-branes. We believe that these issues deserve further
study and that they may provide us with physically important
consequences.

One intriguing model would be a system of static D-branes and a
ghost D-brane moving toward the D-branes. Since the transverse
scalars of the ghost D-brane has the wrong signs in their kinetic
terms, this system would describe a ghost condensation. Also if we
assume that the D-branes are far apart from the ghost D-brane, the
influence of the ghost D-brane on an observer sitting on the
D-branes will be tiny, as the bulk theory is described by the
ordinary superstring theory. Keeping in mind the non-unitarity and
instabilities of the ghost fields, it may still be interesting to explore
applications of ghost branes to various phenomenological
purposes. For example, the possible role of ghost fields (a.k.a. phantom matter) 
as dark energy has been discussed in cosmology \Caldwell. 
Ghosts also appear in \GC. An even closer example may be \KaSu, where the
$Z_2$ symmetry which flips the sign of energy is proposed in the
matter sector as a possible resolution of the cosmological constant
problem.

\subsec{IIA/Heterotic Duality}

It is well-known that type II string theory on K3 is dual to
heterotic string theory on $T^4$ \duality \Senhet \HS . In this
correspondence, the moduli space of $N=(4,4)$ $\hat{c}=2$ conformal
field theory on the K3 surface \eqn\cosetkt {SO(4,20,\IZ)\Bigl\backslash
SO(4,20)\Bigl/ SO(4)\times SO(20).} is
 equivalent to the one for the heterotic
string on $T^4$. Under this duality, a fundamental heterotic string
is mapped to an NS5-brane wrapped on the entire K3 surface in type
IIA string theory.

We expect that a similar duality will also hold for our heterotic
string.  This leads to the conjecture that the $OSp(32+2n|2n)$ heterotic
 string on $T^4$ is equivalent to type IIA string
theory on a certain manifold whose sigma model leads to a
$\hat{c}=2$ SCFT with the following moduli space
\eqn\modufin{SO(4+n,20+n,\IZ) \Bigl\backslash
SO(4+n,20+n)\Bigl/SO(4)\times SO(n,20+n),} as is clear from
\narainl.

Since K3 is known to be the unique compact and simply connected
Ricci flat manifold with complex dimension two, we probably need to
consider a supermanifold with super dimension two (i.e., bosonic dim.
$-$ fermionic dim.=2) in the same sprit as in the heterotic string
side. Notice that the same manifold can occur in the F-theory
description discussed in section 5. Moreover, a supermanifold
version of ALE spaces can naturally arise by considering the T-dual
of the system with NS5-branes and ghost NS5-branes.

\subsec{Other Superalgebras}

It would be interesting to ask if finite dimensional Lie
superalgebras other than $U(N|M)$ and $OSp(N|M)$ can appear in some
open string theory as Chan-Paton matrices. To have such an
interpretation, they need to be realized as subalgebras of the
supermatrices $gl(N|M)$. From the Kac classification \Kac\ of Lie
superalgebras, we  find two families of such superalgebras $Q(N)$
and $P(N)$. These are called strange superalgebras and are
defined by the supermatrices
\eqn\supermatqn{\Phi_{Q(N)}=\left(\matrix{\phi &\psi \cr \psi &\phi
\cr}\right),\ \ \ \ \ \ \ \Phi_{P(N)}=\left(\matrix{\phi &\psi_{S}
\cr \psi_{A} &-\phi^{t} \cr}\right),} where $\phi$ and $\psi$ are
$N\times N$ bosonic and fermionic matrices,\foot{Strictly speaking
the mathematical definition of $Q(N)$ and $P(N)$ requires that
$\phi$ and $\psi$ are traceless to make the algebra simple. We
ignore this condition here as we usually do when discussing the
$U(N)$ gauge symmetry on $N$ D$p$-branes.} respectively;
$\psi_S$ and $\psi_A$ are the fermionic symmetric and antisymmetric
matrices.

Formally, we can find a $\IZ_2$ action $h$ which projects a system
of $N$ D-branes and $N$ ghost D-branes into the system that
corresponds to \supermatqn. They are given\foot{Here we have type II
strings in mind. In the bosonic string case, we can just set
$F_S=0$.} by $h_{Q}=-(-1)^{F_S}$ and $h_{P}=-\Omega(-1)^{F_s}$
($F_S$ is the spacetime fermion number) for $Q(N)$ and $P(N)$
respectively. This is because the action $-(-1)^{F_S}$ flips the
signs of the NSNS and RR parts of a boundary state and a D-brane is
mapped to a ghost D-brane (i.e. it acts as $\Phi\to
\sigma_2\Phi\sigma_2$). However, the closed string theories arising
from projections by $h_Q$ and $h_P$ do not seem to make sense; in
particular the OPEs may not close.

Other interesting superalgebras are the exceptional ones. There are
two of them: $F(4)$ and $G(3)$. Since the non-simply laced Lie group
$F_4$ can be found as part of a gauge group in the CHL string theory
\CHL,  these supergroups might somehow show up in  heterotic string
theory.

\subsec{Future Directions}

There are various `ghost' branes whose existence in string theories
are predicted by dualities.
These include the negative tension versions of NS5-branes, fundamental strings,
M2-branes, and M5-branes.
It would be interesting to study the properties of these objects.

The reduction of a system with supergroup symmetries to
one with usual bosonic symmetries
as described in subsection 3.1 holds not just in
 string theory but in general quantum field theories.
It may be possible to find useful applications of this.

It is known that the perturbation theory does not appear to reduce
to $SU(N-M)$ around a vacuum of the $SU(N|M)$ Chern-Simons theory
\V. More generally, one can consider a topological brane-anti brane
system where some background fields are turned on and anti branes
are not completely canceled. We expect that the topological string
amplitudes for such a system computes some terms in the low-energy
action of corresponding system of D-branes and ghost D-branes, by
replacing traces with supertraces in the usual formulas. This may
have some applications.

\vskip .3in

\noindent
{\it Note added:}
A few weeks after our paper appeared on the web, we received an
interesting paper \EMR, where  ghost D3-branes are applied to
discuss a version of AdS/CFT correspondence. There, the
holographically dual theory is the 4D $N=4$ $SU(N|N)$ super Yang-Mills
theory. As pointed out in that paper, non-superymmetric gauge
theories with the supergroup $SU(N|N)$ had been used to realize
gauge invariant regularization of  $SU(N)$ theories in a series of
work starting with \Morris.

\vskip .5in

\centerline{\bf Acknowledgments} We are grateful to J. Polchinski
and S. Ryu for useful discussions. We also  thank R.
Emparan, S. Giddings, E. Gimon, G. Horowitz, Y. Hyakutake, D. Mateos, J-H.
Park, S. Parkhomenko, M. Shigemori, T.Tokunaga, P. Yu-song for
helpful comments. This work was supported in part by the National
Science Foundation under Grant No.~PHY99-07949 and  PHY04-56556.

\vskip .3in

\appendix{A}{Definition of Lie Superalgebras}

 Lie superalgebras \Kac\ are defined by replacing the commutation
relations in the definition of usual Lie algebras with the $\IZ_2$
graded ones such that (anti-)commutators satisfy
\eqn\comut{[X,Y]=-(-1)^{|X||Y|}[Y,X]; \ \ \ \ \
[X,[Y,Z]]=[[X,Y],Z]+(-1)^{|X||Y|}[Y,[X,Z]],} where $|X|$ denotes the
fermion number of $X$, i.e. $|X|=0$ if $X$ is even (or bosonic) and
$|X|=1$ if $X$ is odd (or fermionic). Finite dimensional Lie
superalgebras were classified by Kac \Kac. Some of them can be
expressed by supermatrices. For more extensive reviews, refer to
\FrappatPB \DeWittCY \FreundWS.

\subsec{Preliminaries}

We express a supermatrix $X$ in terms of bosonic $A,B$ or fermionic
$C,D$ submatrices \eqn\grparassm{X=\left(\matrix{A& B\cr
C&D\cr}\right)\in gl(N|M).} We define supertrace ${\rm Str}$ and
superdeterminant ${\rm Sdet}$ of $X$ by \eqn\supertraces{{\rm Str}X
=\Tr A -\Tr D,\ \ \ \ \ \ {\rm Sdet}X=\det(A-BD^{-1}C)\cdot
\det(D)^{-1}.} They satisfy \eqn\propertysup{\Str(MN)=\Str(NM),\ \
{\rm Sdet}(MN)={\rm Sdet}(M){\rm Sdet}(N),\ \ {\rm Sdet}(e^{M})=
e^{\Str M}.}
In order to be consistent with these, we work with a
supertransposed matrix
$$X^{\ti{T}}=\left(\matrix{A^T&C^T\cr
-B^T&D^T}\right)  $$
 instead of a usual transposed matrix $X^T$. We can
indeed show that \eqn\propsutra{{\rm Str}(X^{\ti{T}})={\rm Str}X,\ \ \ \
{\rm Sdet}(X^{\ti{T}})={\rm Sdet}X,} and
\eqn\protr{(XY)^{\ti{T}}=Y^{\ti{T}}X^{\ti{T}}.}
Note that $\T$ is not a $\IZ_2$ action, but a $\IZ_4$ action.
In fact we can see \eqn\zfour{(X^{\ti{T}})^{\ti{T}}=K X K=
\left(\matrix{A& -B\cr -C&D\cr}\right),} where $K$ denotes
\eqn\kdef{K=\left(\matrix{I_{N}& 0\cr 0&-I_{M}\cr}\right).}

On the other hand, the adjoint operation remains the ordinary one
\eqn\adjo{X^{\dagger}=(X^T)^*,} which
involves  $T$ rather than $\T$. This satisfies
\eqn\prodj{(XY)^{\dagger}=Y^{\dagger}X^{\dagger},\ \ \  {\rm and}\ \
\ \ (X^{\dagger})^{\dagger}=X. } We call a supermatrix $X$ hermitian
if $X^{\dagger}=X$.

\subsec{$SU(N|M)$}

An element of the Lie superalgebra $SU(N|M)$ (also called $A(N-1|M-1)$)
is a hermitian supermatrix $\Phi$ whose supertrace is zero: $\Str\Phi=0$.
 It is divided into bosonic and fermionic elements
\eqn\grparassm{\Phi=\left(\matrix{\phi^1_{ij}&\psi_{ia}\cr
\psi^*_{ia}&\phi^2_{ab}\cr}\right),} where $1\leq i,j\leq N$ and
$N+1\leq a,b\leq N+M$. The matrices $\phi^1$ and $\phi^2$ are
bosonic and belong to the Lie algebras $U(N)$ and $U(M)$, with the
constraint $\sum_{i}\phi^1_{ii} -\sum_{a}\phi^2_{aa}=0$. The complex
matrix $\psi$ is fermionic. Fig.5 is the Dynkin diagram of
$SU(N|M)$.

\fig{The Dynkin diagram of the superalgebra $SU(N|M)$ (also called
$A(N-1|M-1)$). Each node, labeled by an integer $m$, represents a
simple root $\ap_m$ $(m=1,2,\ddd,N+M)$. The gray
node $\otimes$ is a fermionic root $\ap_N$ and its length is zero.
The other roots represented by white nodes
$\bigcirc$ have length-squared $(\ap_m,\ap_m)=\pm
2$.}{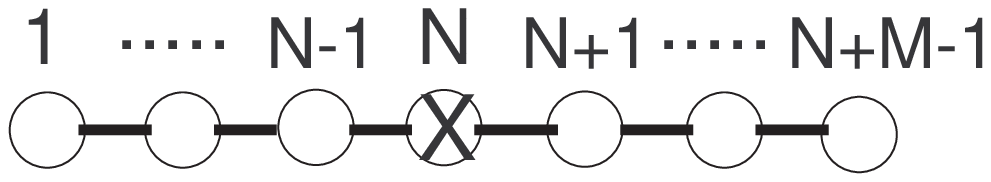}{2.2truein}

In the special case $N=M$ the algebra is not simple
because the element $I_{2N}$ commutes with everything else.
We have to take a quotient by $U(1)$ to make it simple. This is called $PSU(N|N)$.

\subsec{$OSp(N|M)$}

We move on to the orthosymplectic algebra $OSp(N|M)$, where $M$ is always even.
In
Kac's classification, this is called $D(N/2, M/2)$ if $N$ is even,
$C({M\over 2}+1)$ if $N=2$, and $B({N-1\over 2},{M\over 2})$ if $N$ is odd.

 The $OSp(N|M)$ is defined by imposing the
constraint on $SU(N|M)$,  i.e. \eqn\superasym{\gamma\cdot
\Phi+\Phi^{\ti T}\cdot \gamma=0,} where $\gamma$ is defined by
\eqn\gam{\gamma=\left(\matrix{I_N&0\cr 0&\eta_M\cr}\right),} where
$\eta_M$ is the $M\times M$ matrix
\eqn\defj{\eta_M=iJ_{M}=\left(\matrix{0&iI_{M/2}\cr
-iI_{M/2}&0\cr}\right).} Solutions to \superasym\ are the
superalgebra elements of $OSp(N|M)$ and can be written as
\eqn\matospp{\Phi=\left(\matrix{\phi^1_{ij}&-(\psi^T\eta)_{ia}\cr
\psi_{ai}&\phi^2_{ab}\cr}\right).} The first bosonic part $\phi^1$
satisfies the $O(N)$ projection \eqn\probos{(\phi^1)^T=-\phi^1,}
while the second one $\phi^2$ satisfies the $Sp(M)$ (or $USp(M/2)$)
projection \eqn\probosp{(\phi^2)^T=-\eta\cdot \phi^2\cdot \eta.} Due
to the hermiticity condition, the fermionic part obeys
\eqn\fermicon{ \psi^*=\eta\psi.} Fig.6 is the Dynkin Diagram of the
superalgebra $OSp(2M|2N)=D(N,M)$.

\fig{The Dynkin diagram of the superalgebra $OSp(2M|2N)$ (or called
$D(N,M)$). Each node, labeled by an integer $m$, represents a
simple root $\ap_m$ $(m=1,2,\ddd,N+M)$. The gray
node $\otimes$ is a fermionic root $\ap_N$, whose length is zero.
The other simple roots expressed by the while nodes
$\bigcirc$ have  length-squared $(\ap_m,\ap_m)=\pm 2$.  The subdiagram
with the simple roots $\ap_{N+1},\ddd,\ap_{N+M}$ is identical
to the $D_M$ Dynkin diagram.}{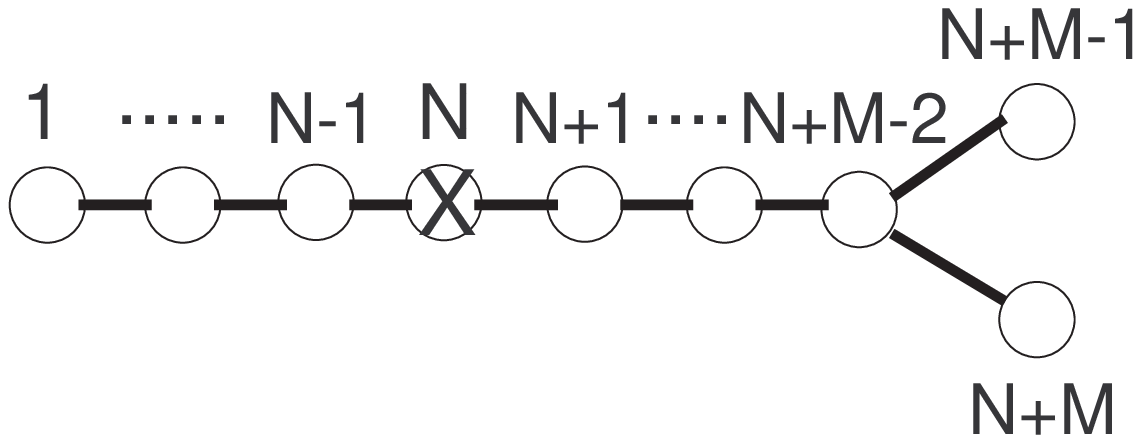}{2.2truein}

\subsec{Other Lie Superalgebras}

There are many other superalgebras in Kac's classification. Here we
summarize them.
In general, Lie superalgebras fall into two classes: classical Lie superalgebras
and Cartan type superalgebras.

In addition to  $SU(N|M)$ and $OSp(N|M)$
(also called $A(n,m),~B(n,m),~C(n+1),~D(n,m)$ in Kac's classification
as we mentioned), the classical Lie superalgebras include the
exceptional ones called $F(4)$ and $G(3)$. They have as the bosonic
parts the Lie algebras of $SO(7)\times SU(2)$ and $G_2\times SU(2)$,
respectively. Also
it is known that $D(2,1)$ has a continuous parameter $\ap$ and is
called $D(2,1;\ap)$. Furthermore, the classical superalgebras also
include the so-called strange superalgebras denoted by $Q(n)$ and
$P(n)$.

Finally, there are four families of Cartan type superalgebras called
$W(n),S(n),\ti{S}(n)$ and $H(n)$. They are defined as (sub)algebras
of the vector fields on the $n$ dimensional flat fermionic manifold,
whose coordinates are given by $n$ Grassmann numbers
$(\theta_1,\theta_2,\ddd,\theta_n)$.

\appendix{B}{Proofs of Cancellation in $U(N|M)$ and $OSp(N|M)$ Super
Matrix Models}
\subsec{Proofs by Virasoro constraints}

The $U(N|M)$ supermatrix model is defined by the action
\eqn\actionmat{ S=-{\rm Str}V(\Phi),~~V(\Phi)=\sum_{n\geq
1}c_n\Phi^n } and the matrix integral \eqn\partss{Z[U(N|M)]=\int d\Phi\
e^{-S(\Phi)}. } We can derive an infinite number of partial
differential equations which should be satisfied by the partition
function $Z[U(N|M)]$ in the form of  Virasoro constraints. They
can be found by shifting the supermatrix as
\eqn\supershift{\Phi\rightarrow \Phi'=\Phi+\epsilon\left({1\over
z-\Phi}\right).} To compute the Jacobian, let us consider the
variation \eqn\sjbo{\eqalign{ \delta \Phi'&=\delta \Phi+\epsilon
\sum_{k=0}^\infty\sum_{l=0}^{k}{1\over z^{k+1}} \Phi^l\delta\Phi
\Phi^{k-l}.}} Then the superJacobian reads \eqn\sj{
J=1+\epsilon\left(\str{1\over z-\Phi}\right)^2.} By combining the
superJacobian and the variation of the action, we obtain the loop
equation
$$
\left\langle \left(\str{1\over z-\Phi}\right)^2+\Str { V'(\Phi)\over
z-\Phi} \right\rangle=0.
$$
The ${\cal O}(1/z^{k+2})$ term reads
$$
\left\langle \sum_{l=0}^k\str\Phi^{l} \Str \Phi^{k-l} +\sum_n n
c_n\Str \Phi^{n+k} \right\rangle=0,
$$
which can also be written as
$$
0=L_k Z\equiv\left(2(N-M)\p_k+ \sum_{l=1}^{k-1}\p_l\p_{k-l}
+\sum_{n=0}^\infty n c_n \p_{k+n} \right)Z.
$$
for $k\geq 1$ and \eqn\erzero{\eqalign{ & 0=L_0 Z\equiv
\left((N-M)^2+\sum_{n=0}^\infty n c_n \p_n\right)Z,\cr &0=L_{-1}Z
\equiv \left(c_1(N-M)+\sum_{n=0}^\infty (n+1) c_{n+1}
\p_n\right)Z.}}
These differential operators satisfy the Virasoro algebra
$$
[L_k,L_l]=(k-l)L_{k+l}.
$$
The fact the Virasoro generators depend on $N$ and $M$ only through
the combination $N-M$ suggests that the dynamics of the supermatrix
model is identical to that of the $U(N-M)$ matrix model.

We move on to the $OSp$ case.
Consider the supermatrix model
\eqn\OSpmat{
Z[OSp(N|M)]=\int d\Phi e^{\Str V(\Phi)},~~ V(\Phi)=\sum_n c_n \Phi^{2n}.
}
The integral is over the Lie superalgebra of $OSp(N|M)$.
We include only even powers of $\Phi$ in $V$ because the supertrace
vanishes on odd powers. Consider
the following infinitesimal variation in the Lie superalgebra
direction:
$$
\Phi\rightarrow \Phi'=\Phi+\epsilon\left({1\over z-\Phi}\right)_{\rm
odd}
=\Phi+\epsilon{\Phi\over z^2-\Phi^2}.
$$
One can show\foot{$\xi^{\mu\nu}:=(\Phi \gamma^{-1})^{\mu\nu}$ are
`anti-symmetric': $\xi^{\nu\mu}=-(-1)^{|\mu||\nu|}\xi^{\mu\nu}$.
This property makes it easy to consider independent components.}
that the superJacobian to order $\epsilon$ is
$$
J= 1+{\epsilon z^2\over
2}\left(\str{1\over z^2-\Phi^2}\right)^2 -{\epsilon\over
2}\str{1\over z^2-\Phi^2}
$$
and we obtain the Ward
identity
$$
\left\langle \left(\str{z\over z^2-\Phi^2}\right)^2 -\str{1\over
z^2-\Phi^2}+2\Str {\Phi V'(\Phi)\over z^2-\Phi^2} \right\rangle=0.
$$
This is equivalent to the Virasoro constraints
$$
0=L_k Z\equiv\left( {1\over 2}(N-M-{1\over 2})\p_k+ {1\over
4}\sum_{l=1}^{k-1}\p_l\p_{k-l} +\sum_{n=0}^\infty n c_n \p_{k+n}
\right)Z.
$$
for $k\geq 1$ and
$$
0=L_0 Z\equiv \left({1\over 4}(N-M)(N-M-1)+\sum_{n=0}^\infty n c_n
\p_n\right)Z.
$$
The appearance $N$ and $M$ through $N-M$
indicates that the $OSp(N|M)$ model reduces to the $SO(N-M)$ model.

\subsec{Perturbative Proof of Cancellation in the $U(N|M)$ Matrix Model}

We consider correlation functions in the $U(N|M)$ supermatrix model
\actionmat. The propagators can be written as (we set $c_2=1$ by
rescaling) \eqn\propagatoru{\eqalign{\la
\phi^1_{ij}\phi^1_{kl}\lb&=\delta_{il}\delta_{jk},\cr \la
\phi^2_{ij}\phi^2_{kl}\lb&=-\delta_{il}\delta_{jk},\cr \la
\psi_{ij}\psi^*_{kl}\lb&=-\delta_{il}\delta_{jk}.}} We can also
write them in a compact way \eqn\superpro{\la
\Phi_{\mu\nu}\Phi_{\rho\sigma}\lb=\delta_{\mu\sigma}
\delta_{\nu\rho}(-1)^{|\nu|}.} In this notation the supertrace is
given by ${\rm Str}\Phi=\sum_{\mu}(-1)^{|\mu|}\Phi_{\mu\mu}$.

Below we follow the arguments in \AM\ to show that any correlation
function in this matrix model only depends on $N-M$. Since we can
perform perturbative expansions of interaction terms, we have only
to examine correlation functions in the free theory (i.e. $c_n=0$
for $n\geq 3$). To make the fermionic nature of the indexes manifest, we
rewrite $\Phi_{\mu\nu}$ as $\ap_{\mu}\bar{\ap}_{\nu}$. Then
$\ap_{i}$ are bosonic while $\ap_{a}$ are fermionic. Consider the
operator product ${\rm Str}\Phi^3{\rm Str}\Phi^3$ which can be
expressed as
\eqn\expone{\eqalign{&(-1)^{|\mu|}\ap_{\mu}\bar{\ap}_{\nu}\ap_{\nu}
\bar{\ap}_{\rho}\ap_{\rho}\bar{\ap}_{\mu}\cdot (-1)^{|\xi|}\ap_{\xi}
\bar{\ap}_{\eta}\ap_{\eta}
\bar{\ap}_{\sigma}\ap_{\sigma}\bar{\ap}_{\xi}\cr &=
(\bar{\ap}_{\mu}\ap_{\mu})(\bar{\ap}_{\nu}\ap_{\nu})
(\bar{\ap}_{\rho}\ap_{\rho})\cdot( \bar{\ap}_{\xi}\ap_{\xi})
(\bar{\ap}_{\eta}\ap_{\eta})(\bar{\ap}_{\sigma}\ap_{\sigma}).}} Then
we take the Wick contractions using the propagator \superpro. We
concentrate on a particular contraction. Then by moving only pairs
of $(\bar{\ap}\ap)$ we can always  divide \expone\ into several
parts such that in each part the contraction is taken successively
following the array of $\ap_\mu$s
\eqn\eprtwo{\eqalign{&\Big[(\bar{\ap}_{\mu}
\widehat{\ap_{\mu})(\bar{\ap}_{\mu'}}\widehat{\ap_{\mu'})\cdot}\ddd\Big]\cdot
\Bigl[(\bar{\ap}_{\nu}\widehat{\ap_{\nu})(\bar{\ap}_{\nu'}}\widehat{
\ap_{\nu'})\cdot}\ddd\Bigl]\cdot \ddd\cr &=(-1)^{\#{\rm fermionic~
loops}}\cdot
\left[\widehat{\ap_{\mu}\bar{\ap}_{\mu'}}\widehat{\ap_{\mu'}\cdot}\ddd
\bar{\ap}_{\mu}\right]\cdot
\left[\widehat{\ap_{\nu'}\bar{\ap}_{\nu}}\widehat{\ap_{\nu}\cdot}
\ddd\bar{\ap}_{\nu'}\right]\cdot \ddd,}} where $\#$fermionic loops
denotes the number of loops where the sum with respect to
$a=N+1,\ddd,N+M$ is taken. The contraction is denoted by
$\widehat{\ap\bar{\ap}}$. Now evaluate the
contraction using the propagator \superpro. We do this as
follows \eqn\comp{\widehat{\ap_{\mu}\bar{\ap_{\lambda}}}
\widehat{\ap_{\rho}\bar{\ap_{\nu}}}=(-1)^{|\nu|}\la
\Phi_{\mu\nu}\Phi_{\rho\lambda}\lb=\delta_{\mu\lambda}\delta_{\nu\rho}.}
 Indeed it is easy to see that
we can simply replace each contraction
$\widehat{\ap_\mu\bar{\ap}_\nu}$ with the $\delta_{\mu\nu}$. In this
way the sum over each Feynman diagram looks like
\eqn\sumober{\eqalign{&\la {\rm Str}\Phi^{n_1}{\rm Str}\Phi^{n_2}\ddd\lb
\cr &=\sum_{{\rm All~
diagrams}}\left[\sum_{\mu_1}(-1)^{|\mu_1|}\delta_{\mu_1\mu_1}\right]
\cdot \left[\sum_{\mu_2}(-1)^{|\mu_2|}\delta_{\mu_2\mu_2}\right]\ddd
\left[\sum_{\mu_L}(-1)^{|\mu_L|}\delta_{\mu_L\mu_L}\right]\cr
&=\sum_{{\rm All~ diagrams} }(N-M)^{L},}} where $L$ is the number of
loops in each Feynman diagram. Thus we have shown \formula. Note
that the dependence on $N$ and $M$ only through $N-M$ holds for individual  Wick
 contractions.
 This shows that the $U(N|M)$ matrix model reduces to the $U(N-M)$ matrix model with
the same action.

So far we have been considering a system with $U(N|M)$ gauge symmetry,
or a system where the fields transform in the adjoint representation of $U(N|M)$.
We can also consider a situation with a $U(N'|M')$ flavor symmetry.
This can be modeled by  bi-fundamental fields $Q_{\mu m}, \bar{Q}_{m \mu}$, where
$\mu$ and $m$ are  vector indices for $U(N|M)$ and $U(N'|M')$, respectively.
The propagator is
$$
\langle Q_{\mu m}\bar{Q}_{n\nu}\rangle=(-1)^{|m|}\delta_{\mu\nu}\delta_{mn}.
$$
By writing
$$
Q_{\mu m}=\beta_\mu \bar{\gamma}_m,~~\bar{Q}_{m\mu}=\gamma_m\bar{\beta}_\mu,
$$
the argument above goes through and shows that the results of perturbative computations
depend on $N'$ and $M'$ only through $N'-M'$.
Thus a system with $N'|M'$ quarks is equivalent to a system with $N'-M'$ quarks.

\subsec{Perturbative Proof of Cancellation in the $OSp(N|M)$ Matrix Model}

Next we consider the perturbative expansion of \OSpmat. The
supermatrix takes the form \matospp. The propagators are given by
(we set $c_2=1/4$) \eqn\propsp{ \eqalign{\la
\phi^1_{ij}\phi^1_{kl}\lb&=\delta_{il}\delta_{jk}-\delta_{ik}\delta_{jl},\cr
\la
\phi^2_{ab}\phi^2_{cd}\lb&=-\delta_{ad}\delta_{bc}-\eta_{ac}\eta_{bd},\cr
\la \psi_{ai}\psi_{bj}\lb&=-\delta_{ij}\eta_{ab},\cr \la
(-\psi^T\eta)_{ia}\psi_{bj}\lb&=-\delta_{ab}\delta_{ij}.}}
We can summarize \propsp\ as  \eqn\coreospps{\la
\Phi_{\mu\nu}\Phi_{\rho\sigma}\lb=(-1)^{|\nu|}\delta_{\mu\sigma}\delta_{\nu\rho}
-(-1)^{|\mu||\nu|}\gamma_{\sigma\nu}\gamma_{\mu\rho},} where
$\gamma_{\mu\nu}$ is defined in \gam. It is easy to see that
\coreospps\ is consistent with the projection \superasym.

To show \formulaosp, let us first compare this
with the previous $U(N|M)$ case. The only difference is the second
term in the propagator \coreospps. We can apply the argument
in the $U(N|M)$ case to contractions involving only the first term.
The total expression of the correlation function can be obtained by
replacing some of the contractions
$(-1)^{|\nu|}\delta_{\mu\sigma}\delta_{\nu\rho}$s with the second
one $-(-1)^{|\mu||\nu|}\gamma_{\sigma\nu}\gamma_{\mu\rho}$s.

Remember that the previous result in the $SU(N|M)$ case holds for each
Wick contraction, which looks like
\eqn\wickcont{\bigl[(-1)^{|\mu_1|}\delta_{\mu_1\mu_2}\delta_{\mu_2\mu_3}
\ddd\delta_{\mu_A\mu_1}\bigl]\cdot
[(-1)^{|\nu_1|}\delta_{\nu_1\nu_2}\delta_{\nu_2\nu_3}
\ddd\delta_{\nu_B\nu_1}\bigl]\cdot \bigl[\cdot\bigl]\ddd.} Let us
replace one of the propagators with the one in the second term.
Without losing generality we can replace
$\delta_{\mu_1\mu_2}\delta_{\nu_1\nu_2}$ with
$-(-1)^{|\mu_2|+|\mu_1||\mu_2|}\gamma_{\mu_2\nu_1}\gamma_{\mu_1\nu_2}$.
Then \wickcont\ is changed into
\eqn\wicknew{-(-1)^{|\mu_1|+|\nu_1|}\cdot
(-1)^{|\mu_2|+|\mu_1||\mu_2|}\gamma_{\mu_2\nu_1}
\gamma_{\mu_1\nu_2}\delta_{\mu_1\mu_2}
\delta_{\nu_1\nu_2}= (-1)^{|\mu_1|}\delta_{\mu_1\mu_1},} where we
have employed the identity
$\gamma_{\mu\nu}=(-1)^{|\mu||\nu|}\gamma_{\nu\mu}$ and $\gamma^2=1$.
Therefore we have shown that this result depends only on $N-M$ again,
though the power of $N-M$ is reduced by one. More general cases can be
handled by induction. This completes the proof of \formulaosp.

\subsec{Anomaly Cancellation}

Here we show that the results \anomalytwo\
can be found by just replacing
$\Tr$ in the results for $O(N-M)$ with the supertrace $\Str$.
When we compute $\Tr_{ad}[F^{2m}]$ by taking only the first term in
\genera\ into account,
then it is obvious that we obtain $\Str[1]\Str[F^{2m}]=(N-M)\Str[F^{2m}]$.
 In order to incorporate
other terms we can replace $\delta_{\lambda\nu} t^A_{\mu\rho}$ with
them. For example, let us start with the desired expression (i.e.
written in terms of supertrace) in a general from
 \eqn\startt{(-1)^{|\lambda|}\delta_{\lambda\nu}
M_{\nu\lambda}(-1)^{\mu}t^A_{\mu\rho}N_{\rho\mu}\ddd,} for some
matrices $M$ and $N$. If we replace
 $\delta_{\lambda\nu} t^A_{\mu\rho}$
with the second term, then we find \eqn\comfu{(-1)^{|\lambda|}
M_{\nu\lambda}(-1)^{\mu}N_{\rho\mu}
(-1)^{(|\mu|+|\lambda|)(|\lambda|+|\nu|)+|\lambda|+|\mu|}
\delta_{\mu\rho}t^A_{\lambda\nu} =-(-1)^{|\mu|}N_{\mu\mu}\cdot
(-1)^{|\lambda|} t^A_{\lambda\nu}M_{\nu\lambda}.} Therefore again we
can find the desired form. For other two terms, we can proceed in
the same
 way remembering the identities used in \wicknew.
In this way  \anomalytwo\ is proved by induction.

\appendix{C}{Some Properties of
Affine Super Algebras}

A Lie superalgebra has its affine extension just like an ordinary
Lie algebra. The affine super Kac-Moody algebras at level-$k$ is
defined by \currentalsu\ with $\Str[t^At^B]$ replaced by
$k\Str[t^At^B]$. It defines a conformal field theory\foot{Notice
that in these supergroup cases, sometimes it is possible to have
conformal field theories even if we do not turn on the WZW
interaction terms. Such a model is called the principal chiral model
and it is indeed conformal when $G=PSU(N|N)$ and $G=OSp(M+2|M)$ i.e.
when $h$ vanishes \V \Ber. Such a model may also be relevant to
$N=2$ string theory as a holographic description.} via the Sugawara
construction \GOW \BCMN.

In general the central charge of the
corresponding Virasoro algebra (obtained from the Sugawara
construction) for the Lie supergroup $G$ is given by
\eqn\centra{c=\f{k \cdot {\rm sdim}G}{k+h},} where $k$ is the level
and sdim$G=$dim$G_{B}-$dim$G_{F}$ is the super dimension of the
supergroup $G$. $h$ is the dual Coxeter number.

In the affine $SU(N|M)$ algebra case, we find \eqn\sdimu{{\rm
sdim}G=(N^2+M^2-1)-2NM=(N-M+1)(N-M-1),} and \eqn\chsi{h=N-M.}
Therefore its central charge is given by
\eqn\cenu{c=\f{k(N-M+1)(N-M-1)}{k+N-M}.}  This only depends on the
difference $N-M$ as expected. Notice that at $k=1$ it
leads to $c=N-M-1$ as can be understood from the vertex operator
construction.

In the $OSp(N|M)$ case, we obtain  \eqn\sdimos{{\rm
sdim}G=\f{N(N-1)}{2}+\f{M(M+1)}{2}-NM =\f{1}{2}(N-M)(N-M-1),} and
\eqn\chsip{h=N-M-2.} Therefore the central charge is given by
\eqn\censp{c=\f{k(N-M)(N-M-1)}{2(k+N-M-2)}.} It again only depends
on the difference $N-M$ as expected. At $k=1$ we find
$c=\f{N-M}{2}$. For example, the affine CFT $OSp(32+2n|2n)$ , the
central charge remains the same as $SO(32)$. In particular it is
$c=16$ at level one $k=1$.

\appendix{D}{Momenta in Toroidal Compactification of
the $OSp(32+2n|2n)$ or $E(8+{n\over 2},{n\over 2})\times E(8+{n\over 2},{n\over 2}) $ Heterotic string}
Here we derive the formulas \momlattice\ for the momenta
by generalizing the arguments in \refs{\NSW}.

We consider the compactification of the $OSp(32+2n|2n)$ or $E(8+{n\over 2},{n\over 2})\times E(8+{n\over 2},{n\over 2}) $
heterotic
string on $T^d$ and turn on metric, $B$-field, and Wilson lines as
in subsection 4.5. The world-sheet theory contains free bosons
$X^{\mu=1,\ddd,d}$ with radius $R$, left-moving free bosons
$X^{k=n+1,\ddd,16+2n}, X^{l=1,\ddd,n}$ with radius
$\sqrt{2\alpha'}$.\foot{ In the notation of  subsection 4.3,
$X^k=\sqrt{\alpha'\over 2} \varphi^{k-n},~~X^l=\sqrt{\alpha' \over
2}\tilde{\varphi}^l.$ } There are also $n$ pairs of $\eta\xi$
fermions of conformal weights 1 and 0. We can ignore these fermions
here. The world-sheet theory is described by the following
world-sheet action:
$$\eqalign{
S&={1\over 4\pi}\int d\tau \int_0^{2\pi} d\sigma
\left[
{1\over\alpha'}(g_{\mu\nu}\p_\alpha X^{\mu} \p^\alpha X^{\nu}-
\epsilon^{\alpha\beta} B_{\mu\nu}
\p_\alpha X^{\mu}\p_\beta X^{\nu}\right.\cr
&~~~\left.
+\p_\alpha X^{k}\p^\alpha X^{k}
-\p_\alpha X^{l}\p^\alpha X^{l}) - A^{k}_{\mu}
\epsilon^{\alpha\beta}\p_\alpha X^{k} \p_\beta X^{\mu}
+ A^{l}_{\mu}\epsilon^{\alpha\beta}\p_\alpha X^{l} \p_\beta X^{\mu}
\right]
}
$$
together with the constraints $(\p_\tau-\p_\sigma)X^{k}=0$ and
$(\p_\tau-\p_\sigma)X^{l}=0$.
Here $\epsilon^{\tau\sigma}=-\epsilon^{\sigma\tau}=1$.
We take the radius for $X^{\mu}$ to be $R$.
The radius for $X^{k}$ and $X^{l}$ is $\sqrt{2\alpha'}$, i.e.,  the free fermion radius.
$X^{l}$ are the free bosons that bosonize the spin-${1\over 2}$ $\beta$-$\gamma$ systems,
and appear in the kinetic terms with the wrong sign.

The constraints are second class.
To canonically quantize the system, we need to use the Dirac bracket
to take the constraints into account.
The canonical momenta are
$$\eqalign{
P_{\mu}(\sigma)&={1\over 2\pi\alpha'}
(g_{\mu\nu}\dot{X}^{\nu} - B_{\mu\nu}\partial_\sigma X^{\nu}
+A^{k}_\mu\partial_\sigma X^{k}-A^{l}_\mu\partial_\sigma X^{l}),\cr
P_{k}(\sigma)&={1\over 2\pi\alpha'}
(\dot{X}^{k}-A^{k}_\mu\partial_\sigma X^{\mu}),\cr
P_{l}(\sigma)&={1\over 2\pi\alpha'}(- \dot{X}^{l}+A^{l}_\mu\partial_\sigma X^{\mu}).
}$$
The Dirac brackets among them turn out to be
$$\eqalign{
\{P_{\mu}(\sigma),P_{\mu'}(\sigma')\}_{\rm DB}&=
{1\over 4\pi\alpha'}(A^k_{\mu} A^k_{\mu'}-A^l_{\mu} A^l_{\mu'})
\partial_\sigma\delta(\sigma-\sigma'),\cr
\{P_{\mu}(\sigma),P_{k}(\sigma')\}_{\rm DB}&
={1\over 4\pi\alpha'} A^k_{\mu} \partial_\sigma\delta(\sigma-\sigma'),\cr
\{P_{\mu}(\sigma),P_{l}(\sigma')\}_{\rm DB}&
=-{1\over 4\pi\alpha'} A^l_{\mu} \partial_\sigma\delta(\sigma-\sigma'),\cr
\{P_{k}(\sigma),P_{k'}(\sigma')\}_{\rm DB}&
={\delta_{k k'}\over 4\pi\alpha'} \partial_\sigma\delta(\sigma-\sigma'),\cr
\{P_{l}(\sigma),P_{l'}(\sigma')\}_{\rm DB}&
=-{\delta_{ll'}\over 4\pi\alpha'} \partial_\sigma\delta(\sigma-\sigma'),\cr
\{P_{k}(\sigma),P_{l}(\sigma')\}_{\rm DB}&=0.
}$$
The Dirac brackets involving $X^{\mu}$ are equal to the Poisson brackets.
If we take the combination
$$
P'_{\mu}(\sigma)=P_{\mu}(\sigma)-A^{k}_\mu P_{k}(\sigma)-A^{l}_\mu P_{l}(\sigma),
$$
$P'_{\mu}, P_{k},$ and $P_{l}$ commute among
themselves and are the momenta that are
truly canonically conjugate to $X^{\mu}, X^{k}$,
and $X^{l}$ in the presence of the constraints.
Let $x^{\mu}, x^k,$ and $x^l$ be the zero-modes
of $X^{\mu}, X^{k},$ and $X^{l}$.
The momenta canonically conjugate the zero-modes
are quantized in units of the inverse radii.
Let us write
$$
\int P'_{\mu} d\sigma ={n_{\mu}\over R},
~~\int P_{k} d\sigma={2 q_k\over \sqrt{2\alpha'}},
~~ \int P_{l} d\sigma={2q_l\over \sqrt{2\alpha'}}.
$$
Then $n_{\mu}$ is an integer while $q_k$ and $q_l$ are half integers taking
values in the appropriate lattice defining the heterotic string.
Let $w^{\mu}$ be the winding numbers for $X^{\mu}$.
Then one finds
$$\eqalign{
X^{\mu}(\tau,\sigma)&=x^{\mu}+\alpha' g^{\mu\nu}\left[
{n_\nu\over R}+B_{\nu\rho} R w^\rho-q_k A^{k}_\nu -q_l
A^{l}_\nu-{w^\rho\over 2} (A^{k}_\rho A^{k}_\nu-
A^{l}_\rho A^{l}_\nu)
\right]\tau\cr
&~~~~~~~~~~~+ Rw^{\mu}\sigma+{\rm oscillators},\cr
X^{k}(\tau,\sigma)&=x^k+\sqrt{\alpha'\over 2}
(q_k+A^{k}_\mu R w^{\mu})(\tau+\sigma)+{\rm oscillators},\cr
X^{l}(\tau,\sigma)&=x^l+\sqrt{\alpha'\over 2}
(-q_l+A^{l}_\mu R w^{\mu})(\tau+\sigma)+{\rm oscillators}.
}$$
From this, one can read off the momentum lattice \momlattice.

\listrefs

\end